 \documentclass[journal]{IEEEtran}

\usepackage{amssymb}

\usepackage{amsmath,graphicx,amssymb,bm,textcomp}
\usepackage{hyperref}
\usepackage{cite}
\usepackage{color}
\usepackage{multirow,multicol}
\usepackage{tikz,rotating}
\usepackage{pifont}
\usepackage{paralist}
\usepackage{caption}
\usepackage{stfloats}
\usepackage{tabularx,ragged2e}
\usepackage{balance}

\usetikzlibrary{fit,calc}

\def\layersep{2.5cm}
\usepackage{xcolor}

\definecolor{usc}{rgb}{0.6,0.106,0.117}

\newcommand{\Pg}[1]{\noindent{{\color{red}\textbf{#1}}}}
\renewcommand{\Pg}[1]{}

\newcommand{\eg}{\textit{e}.\textit{g}. }

\newcommand{\subheading}[1]{\noindent\emph{\textbf{\color{black}#1}}}
\newcommand{\cmark}{\ding{51}}%
\newcommand{\xmark}{\ding{55}}%

\newcommand\citet{\cite}

\begin{document}




\title{Transfer Learning from Adult to Children for Speech Recognition: Evaluation, Analysis and Recommendations}


\author{Prashanth Gurunath Shivakumar,~\IEEEmembership{Member,~IEEE} and Panayiotis Georgiou,~\IEEEmembership{Senior Member,~IEEE}%
\thanks{The authors are with the University of Southern California, Los Angeles, USA; e-mail:pgurunat@usc.edu and georgiou@sipi.usc.edu}}

\maketitle

\begin{abstract}
Children speech recognition is challenging mainly due to the inherent high variability in children's physical and articulatory characteristics and expressions. This variability manifests in both acoustic constructs and linguistic usage due to the rapidly changing developmental stage in children's life. Part of the challenge is due to the lack of large amounts of available children speech data for efficient modeling. 
This work attempts to address the key challenges using transfer learning from adult's models to children's models in a Deep Neural Network (DNN) framework for children's Automatic Speech Recognition (ASR) task evaluating on multiple children's speech corpora with a large vocabulary.
The paper presents a systematic and an extensive analysis of the proposed transfer learning technique considering the key factors affecting children's speech recognition from prior literature.
\emph{Evaluations} are presented on (i) comparisons of earlier GMM-HMM and the newer DNN Models, (ii) effectiveness of standard adaptation techniques versus transfer learning, (iii) various adaptation configurations in tackling the variabilities present in children speech, in terms of (a) acoustic spectral variability, and (b) pronunciation variability and linguistic constraints.
Our \emph{Analysis} spans over (i) number of DNN model parameters (for adaptation), (ii) amount of adaptation data, (iii) ages of children, (iv) age dependent-independent adaptation. 
Finally, we provide \emph{Recommendations} on (i) the favorable strategies over various aforementioned - analyzed parameters, and (ii) potential future research directions and relevant challenges/problems persisting in DNN based ASR for children's speech.

\end{abstract}

\begin{IEEEkeywords}
Automatic Speech Recognition, Deep Learning, Transfer Learning, Deep Neural Network, Children Speech Recognition



\end{IEEEkeywords}




\section{Introduction}
\label{sec:intro}
Speech recognition has become an ubiquitous part of our life.
A range of applications, such as human-machine interaction, communication, education, pronunciation and communication tutoring, entertainment and interactive gaming depend on such functionality.
This has become partly possible due to high accuracies achieved by state-of-the-art speech recognition systems.
An important user population for many such technologies are children.
However, Automatic Speech Recognition (ASR) for children is still significantly less accurate than that of adults.
With the recent increased deployment of speech based technologies it becomes ever more important to be inclusive towards children.
Thus there is a need to robustly address the challenges brought by the variability in kids speech.

Researchers have studied how the speech patterns of children differ to that of the adults.
Prior studies have looked into the factors affecting, and degrading, the performance of ASR.
Children speech was found to exhibit high level of variability.
The research suggests that the variability exists in two levels.
Firstly, the variability is embedded in the acoustic signals in the form of spectral and temporal variability, due to the physiological and developmental differences of children.
Secondly, there is variability in kids pronunciation patterns, due to differing and partial linguistic knowledge.

\Pg{Acoustic variability factors:}
Acoustic variability can be attributed to three main factors
(i) shifted overall spectral content and formant frequencies for children \cite{potamianos2003robust},
(ii) high within-subject variability in the spectral content, that affects formant locations \cite{lee1999acoustics},
(iii) high inter-speaker variability observed across age groups, due to developmental changes, especially vocal tract \cite{gerosa2006acoustic}.
\citet{lee1999acoustics} conducted a detailed study analyzing the temporal and spectral parameters of children speech.
The study found that the within-subject variability decreased with increase in age from 5 years to 12 years, reaching adult levels at an age of 15.

\Pg{Impact on speech systems:}
The \emph{word error rates} (WER) for children's ASR were found to be 2 to 5 times worse than adults \cite{potamianos2003robust}.
Due to the specificity associated with children's speech, training children-specific ASR models was found to be highly advantageous.
Age dependent ASR models were also studied giving promising improvements, thereby confirming high inter-age dependent acoustic variability in children \cite{potamianos1997automatic}.
\citet{li2002analysis} studied the effect of speech bandwidth on recognition accuracy.
The study found that the recognition performance degraded more rapidly for children when the bandwidth was reduced from 4kHz to 1.5kHz.
Investigation of the possible causes showed that the average formant frequencies F1, F2 and F3 for children exceeded those of adults by more than 60\% \cite{russell2001automatic}.

\Pg{Prior Work --  Acoustic features:}
Several techniques to tackle the acoustic variability were proposed in recent times.
Different front-end robust features such as \emph{Mel-Frequency Cepstral Coefficients} (MFCC), \emph{Perceptual Linear Prediction} (PLP) cepstral coefficients, and spectrum based filter bank features have been tried \cite{shivakumar2014improving}.
Several minor alterations of front-end features have also been investigated \cite{shivakumar2014improving,umesh2007study, russell2001automatic, ghai2015pitch, shahnawazuddin2016pitch}.
However,  MFCC features have dominated due to their robustness and compatibility with adult ASR systems.

\Pg{Prior Work --  Acoustic adaptation:}
\citet{potamianos2003robust} proposed several front-end frequency warping techniques and speaker normalization techniques with evaluations over different age groups.
Particularly, \emph{Vocal Tract Length Normalization} (VTLN) technique to suppress acoustic variability introduced by the developing vocal tracts in children has become a standard in children ASR systems \cite{shivakumar2014improving, giuliani2003investigating, stemmer2003acoustic}, effectively reducing inter-speaker and inter-age-group acoustic variability.
Adapting acoustic models with \emph{Maximum Likelihood Linear Regression} (MLLR) and \emph{Maximum A-Posteriori} (MAP) was found to be effective \cite{elenius2005adaptation, shivakumar2014improving, gray2014child}.
Further modest gains were achieved using \emph{Speaker Adaptive Training} (SAT) based on \emph{Constrained MLLR} (CMLLR) for children ASR \cite{gray2014child, shivakumar2014improving}.

\Pg{Prior Work --  Pronunciation variability:}
Some research efforts have also concentrated on dealing with the increased pronunciation variability and mispronunciations present in kids due to limited and developing linguistic knowledge.
Performance gap between spontaneous speech recognition and read speech is particularly large for children \cite{gerosa2009review}.
\citet{gerosa2006acoustic} showed that spontaneous speech annotations are extremely useful.
They showed that  language usage efficiency increases with age for children reaching adult levels at 13 years of age i.e., disfluencies decrease with age.
\citet{potamianos1998spoken} performed an  in-depth analysis of linguistic variability in the context of spoken dialogue systems for children.
Inter-speaker linguistic variability was found to be twice the intra-speaker variability.
Mispronunciations in children were found to be twice as high for children of 8-10 years compared to that of 11-14 years, while the trend was reversed for filler pauses.
Age dependencies were also found for the frequency of false-starts, duration, utterance length and breathing.

\Pg{Prior Work --  Language adaptation:}
\citet{das1998improvements} showed that language models trained on children speech were advantageous to using adult models suggesting children use different grammatical constructs. In \cite{gray2014child}, language model adaptation from adult to children showed improvements.

\Pg{Prior Work -- Pronunciation adaptation:}
Children tend to also mispronounce, thus customized dictionaries for children can provide performance benefits \cite{li2002analysis}.
Pronunciation variations among children vary with age. Data-driven pronunciation variation modeling is shown to be useful across children of all ages \cite{shivakumar2014improving}.
However, part of the variations are attributed towards the phonological processes and hence the customization of dictionaries have their limitations \cite{fringi2015evidence}.

\Pg{Prior Work -- Kids Learning Apps:}
There are also significant efforts in speech applications for kids towards learning.
For example \cite{Tepperman2011Agenerativestudentmodel, tulsiani2017acoustic} focused on read speech assessment.
Further, \citet{tong2017multi} focused on pronunciation assessment in Mandarin.
\citet{hagen2007highly} proposed subword unit based speech recognition for children enabling assessment of children speech at finer details and detection of speech events such as partial words and mispronunciations.

\Pg{Prior Work -- DNN efforts:}
More modern methods, specifically related to deep learning, have been extremely successful in improving ASR performance.
The successes of Deep Neural Networks (DNN) have been attributed to DNN's ability to use vast amount of training data and to better approximate the non-linear functions needed to model speech, thus surpassing GMM based ASR systems.
However, relatively less work has investigated DNNs for children's speech probably due to lack of large amounts of children's training data.
\cite{giuliani2015large, cosi2015kaldi} conducted  ASR experiments using a hybrid DNN-HMM based ASR system.
They trained on approximately 10 hours of Italian children's speech giving small improvements over traditional GMM based systems.
\citet{serizel2014vocal} used a DNN to predict the frequency warping factors for VTLN which was later used to train a hybrid DNN-HMM system.
\cite{liao2015large} employed  convolutional long short-term memory recurrent neural networks  to train children ASR for use with Youtube Kids.
They further employed data augmentation through artificially adding noise  for more robustness.
Combining adults' speech with children's speech for training improved results for both adults and children \cite{qian2016mismatched,liao2015large,fainberg2016improving,qian2016improving}.
Particularly, combining  female adult speech in the training was shown to be more advantageous \cite{qian2016mismatched}.
Multi-task learning frameworks for adapting adults' speech to children's speech were presented in \cite{tong2017transfer, tong2017multi}.
In \cite{serizel2014deep,serizel2017deep}, a technique similar to \cite{tong2017transfer} was adopted to overcome limited training data for DNN. 
Most recently, multi-lingual data adaptation in a transfer learning and multi-task learning framework was found to be useful for the task of ASR for children speaking in non-native language \cite{matassoni2018}.

However, most of the prior works pertaining to analysis of children's speech in context of speech recognition has been on gaussian mixture based hidden markov models (GMM-HMM).
Although there has been a wide consensus in the community about the advantages of DNN acoustic modeling for children's speech \cite{qian2016mismatched,liao2015large,fainberg2016improving,qian2016improving,tong2017transfer, tong2017multi,serizel2014deep,serizel2014vocal,giuliani2015large, cosi2015kaldi}, there has been no work to the best of our knowledge, which attempts to evaluate and analyze where the strengths of the DNNs lie in context to children's ASR.
More importantly there is a need for an analysis of the shortcomings of the DNN based ASRs, i.e., problems and challenges persisting in children speech recognition using state-of-the-art speech recognition systems.
Our study attempts to contribute to this gap and provide insights towards future developments.

In this work, we conduct \textbf{\textit{Evaluations}} on large vocabulary continuous speech recognition (LVCSR) for children, to:
\begin{enumerate}
\item Compare older GMM-HMM models and newer DNN models.
\item Investigate different transfer learning adaptation techniques.
  Particularly we look at two factors degrading children ASR: acoustic variability and pronunciation variability in a DNN setup.
\item Assess effectiveness of different speaker normalization and adaptation techniques like VTLN, fMLLR, i-vector based adaptation versus the employed transfer learning technique.
\end{enumerate}

Further, we conduct \textbf{\textit{Analysis}} over the following parameters in context of transfer learning:
\begin{enumerate}
\item DNN model parameters.
\item Amount of adaptation data.
\item Effect of children's ages.
\item Age dependent transformations obtained from transfer learning and their validity, portability over the children's age span.
\end{enumerate}

\textbf{\textit{Recommendations}} are provided from the insights gained from conducting the aforementioned evaluations and analysis for:
\begin{enumerate}
\item Favorable transfer learning adaptation strategies for low data and high data scenarios.
\item Suggested transfer learning adaptation techniques for children of different ages.
\item Amount of adaptation data required for efficient performance over children's ages.

\begin{figure*}[t]
\centering
\begin{turn}{90}
\begin{tikzpicture}[scale=0.4,every node/.style={scale=0.5},every node/.append style={anchor=center},shorten >=1pt,->,draw=black!50, node distance=\layersep,line width=0.1pt]
    \tikzstyle{every pin edge}=[<-,shorten <=1pt]
    \tikzstyle{neuron}=[circle,fill=black!25,minimum size=17pt,inner sep=0pt]
    \tikzstyle{input neuron}=[neuron, fill=green!50];
		\tikzstyle{ivector neuron}=[neuron, fill=black!50];
    \tikzstyle{output neuron}=[neuron, fill=red!50];
    \tikzstyle{hidden neuron}=[neuron, fill=blue!50];
    \tikzstyle{annot} = [text width=8em, text centered];
		\tikzstyle{surround} = [color=green!80, ultra thick, dashed, rounded corners];
		
		\node at (1*\layersep,1.5) {$\vdots$};
		\node at (1*\layersep,-3.5) {$\vdots$};
		\node at (1*\layersep,-11.5) {$\vdots$};
		\node at (1*\layersep,-15.5) {$\vdots$};
		\node at (2*\layersep,-11.0) {$\vdots$};
		\node at (3*\layersep,-11.0) {$\vdots$};
		\node at (5*\layersep,-11.0) {$\vdots$};

		\begin{scope}[local bounding box=softmax1]
    \foreach \name / \y in {1,...,4}
		{	\pgfmathparse{\y<4 ? int(round(\y)) : "40"}
			\path[yshift=5.5cm]node[input neuron, pin=left:\rotatebox{270}{$x_{\pgfmathresult}$}] (I-\name) at (1*\layersep,{-\y-div(\y,4)}) {};
		}
		\foreach \name / \y in {5,...,8}
		{	\pgfmathparse{\y<8 ? int(round(\y-4)) : "100"}
			\path[yshift=5.5cm]node[ivector neuron, pin=left:\rotatebox{270}{$i_{\pgfmathresult}$}] (I-\name) at (1*\layersep,{-\y-div(\y,8)-1}) {};
		}
		\end{scope}
		
		\begin{scope}[local bounding box=softmax2]
    \foreach \name / \y in {9,...,12}
		{	\pgfmathparse{\y<12 ? int(round(\y-8)) : "40"}
			\path[yshift=6.5cm]node[input neuron, pin=left:\rotatebox{270}{$x_{\pgfmathresult}$}] (I_2-\name) at (1*\layersep,{-\y-div(\y,12)-6}) {};
		}
		\foreach \name / \y in {13,...,16}
		{	\pgfmathparse{\y<16 ? int(round(\y-12)) : "100"}
			\path[yshift=6.5cm]node[ivector neuron, pin=left:\rotatebox{270}{$i_{\pgfmathresult}$}] (I_2-\name) at (1*\layersep,{-\y-div(\y,16)-7}) {};
		}
		\end{scope}
    		
		\foreach \name / \y in {1,...,12}
			\path[yshift=1cm]node[hidden neuron] (H_1-\name) at (2*\layersep,{-\y-div(\y,12)}) {};
			
		\foreach \name / \y in {1,...,12}
			\path[yshift=1cm]node[hidden neuron] (H_2-\name) at (3*\layersep,{-\y-div(\y,12)}) {};
			
		\foreach \name / \y in {1,...,12}
			\path[yshift=1cm]node[hidden neuron] (H_3-\name) at (4*\layersep,{-\y-div(\y,12)}) {};
			
		\foreach \name / \y in {1,...,12}
			\path[yshift=1cm]node[hidden neuron] (H_4-\name) at (5*\layersep,{-\y-div(\y,12)}) {};

		\foreach \name / \y in {1,...,8}
		{	\pgfmathparse{\y<8 ? \y : "n"}
			\path[yshift=-1cm]node[output neuron,pin={[pin edge={->}]right:\rotatebox{270}{$y_{\pgfmathresult}$}}] (O-\name) at (6*\layersep,{-\y-div(\y,8)}) {};
		}


		\foreach \source in {1,...,8}
        \foreach \dest in {1,...,12}
            \path (I-\source) edge (H_1-\dest);
						
		\foreach \source in {9,...,16}
        \foreach \dest in {1,...,12}
            \path (I_2-\source) edge[line width=0.05pt] (H_1-\dest);
		\foreach \source in {1,...,12}
        \foreach \dest in {1,...,12}
            \path (H_1-\source) edge (H_2-\dest);
		\foreach \source in {1,...,12}
        \foreach \dest in {1,...,12}
            \path (H_2-\source) edge (H_3-\dest);
		\foreach \source in {1,...,12}
        \foreach \dest in {1,...,12}
            \path (H_3-\source) edge (H_4-\dest);
		\foreach \source in {1,...,12}
			\foreach \dest in {1,...,8}
				\path (H_4-\source) edge (O-\dest);

  	\node[annot,rotate=-90,below of=I-4, node distance=2cm] (il2) {Input layer};
		\node[annot,rotate=-90,below of=I_2-12, node distance=2cm] (il2) {Input layer};
    \node[annot,rotate=-90,above of=O-5] {Output layer};
		
		\draw[surround] ($(softmax1.north west) + (-2.5pt, 15.5pt)$) rectangle
        ($(softmax1.south east) + (15.5pt, -15.5pt)$);
		\draw[surround] ($(softmax2.north west) + (-2.5pt, 15.5pt)$) rectangle
        ($(softmax2.south east) + (15.5pt, -15.5pt)$);
	
\end{tikzpicture}
\end{turn}
\captionsetup{justification=centering}
\caption{Acoustic Variability Modeling \\ \small{Neuron color scheme: Red-Output, Blue-Hidden, Gray-ivector input, Green-MFCC input}}
\label{fig:av}
\end{figure*}
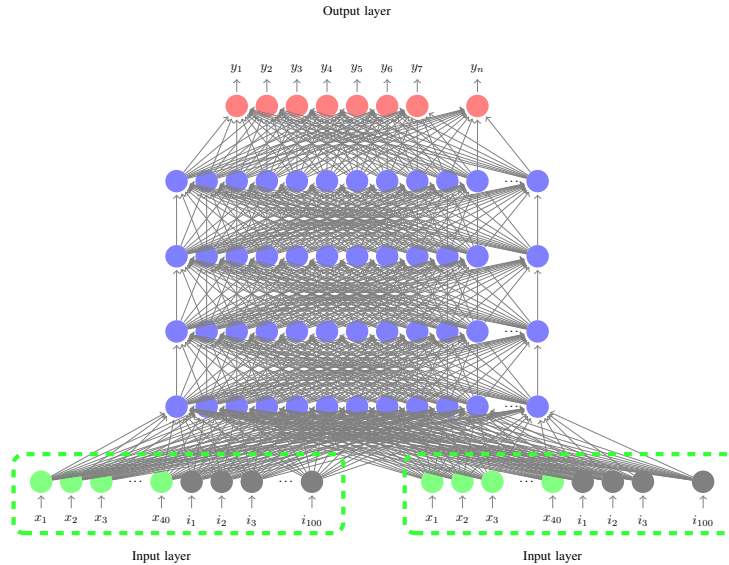

\item potential future research directions and relevant challenges and problems persisting in children speech recognition.
\end{enumerate}

The rest of the paper is formatted as follows:
Section~\ref{sec:proposed} motivates and describes the proposed transfer learning technique.
Section~\ref{sec:database} describes the databases used for recognition experiments.
The experimental setup and baseline systems for both adult and children ASR models are described in Section~\ref{sec:baseline_setup}.
Section~\ref{sec:results} presents experiment results and discussion.
Section~\ref{sec:adaptation_data} analyzes the amount of adaptation data and its effect on the performance.
We carry out analysis of transfer learning technique on children's age in Section~\ref{sec:age_analysis}.
Section~\ref{sec:age_dependent} discusses the study of age dependent transfer learning transformations and Section~\ref{sec:age_indep_vs_dep} provides comparisons between the age dependent and age independent transfer learning transformations.
Finally, Section~\ref{sec:conclusion} discusses potential future work and concludes.

\section{Proposed Transfer Learning Technique}\label{sec:proposed}
\Pg{Transfer Learning Motivation:}
Transfer learning is a method of seeding models of a new task by using the knowledge gained from a related task.
The method has been used successfully, for cross-lingual knowledge transfer in DNN-based speech recognition \cite{heigold2013multilingual, huang2013cross} and character recognition tasks \cite{cirecsan2012transfer}.
Transfer learning often exploits the various level of information that are captured by the different neural network layers.
Often layers closer to the signal capture signal specific characteristics, \eg edge characteristics, basic shapes, or spectral content.
Higher layers capture information more related to the task at hand, \eg phoneme classes, object types \cite{bengio2013representation}.
Children, as described above, differ (i) in acoustics and (ii) pronunciation from adults.
This motivates us to investigate the transfer learning between adult and children ASR systems in two ways:
(i) acoustic variability, as those relate to layers near input, and
(ii) pronunciation variability as it relates to layers near output.




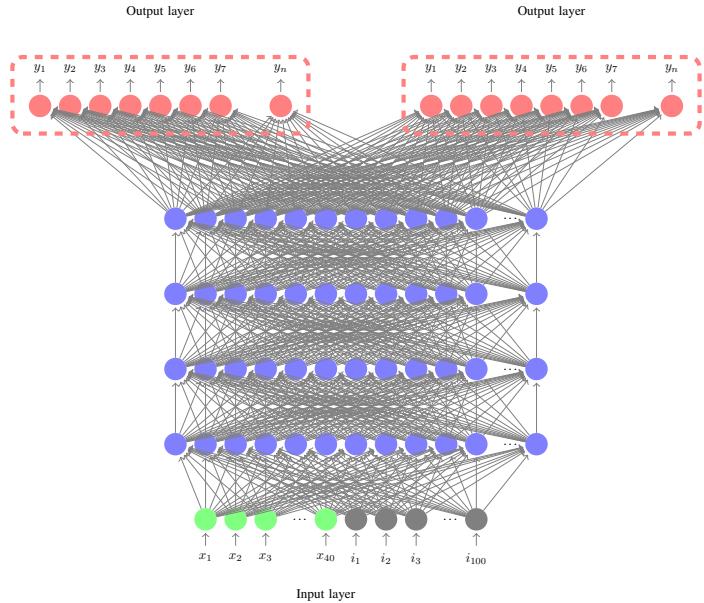
\begin{figure*}[t]
\centering
\begin{turn}{90}
\begin{tikzpicture}[scale=0.4,every node/.style={scale=0.5},every node/.append style={anchor=center},shorten >=1pt,->,draw=black!50, node distance=\layersep,line width=0.1pt]
    \tikzstyle{every pin edge}=[<-,shorten <=1pt]
    \tikzstyle{neuron}=[circle,fill=black!25,minimum size=17pt,inner sep=0pt]
    \tikzstyle{input neuron}=[neuron, fill=green!50];
		\tikzstyle{ivector neuron}=[neuron, fill=black!50];
    \tikzstyle{output neuron}=[neuron, fill=red!50];
    \tikzstyle{hidden neuron}=[neuron, fill=blue!50];
    \tikzstyle{annot} = [text width=8em, text centered];
		\tikzstyle{surround} = [color=red!50, ultra thick, dashed, rounded corners];
		
		\node at (1*\layersep,-4.0) {$\vdots$};
		\node at (1*\layersep,-9.0) {$\vdots$};
		\node at (2*\layersep,-11.0) {$\vdots$};
		\node at (3*\layersep,-11.0) {$\vdots$};
		\node at (5*\layersep,-11.0) {$\vdots$};

    \foreach \name / \y in {1,...,4}
		{	\pgfmathparse{\y<4 ? int(round(\y)) : "40"}
			\path[yshift=0cm]node[input neuron, pin=left:\rotatebox{270}{$x_{\pgfmathresult}$}] (I-\name) at (1*\layersep,{-\y-div(\y,4)}) {};
		}
		\foreach \name / \y in {5,...,8}
		{	\pgfmathparse{\y<8 ? int(round(\y-4)) : "100"}
			\path[yshift=0cm]node[ivector neuron, pin=left:\rotatebox{270}{$i_{\pgfmathresult}$}] (I-\name) at (1*\layersep,{-\y-div(\y,8)-1}) {};
		}
    		
		\foreach \name / \y in {1,...,12}
			\path[yshift=1cm]node[hidden neuron] (H_1-\name) at (2*\layersep,{-\y-div(\y,12)}) {};
			
		\foreach \name / \y in {1,...,12}
			\path[yshift=1cm]node[hidden neuron] (H_2-\name) at (3*\layersep,{-\y-div(\y,12)}) {};
			
		\foreach \name / \y in {1,...,12}
			\path[yshift=1cm]node[hidden neuron] (H_3-\name) at (4*\layersep,{-\y-div(\y,12)}) {};
			
		\foreach \name / \y in {1,...,12}
			\path[yshift=1cm]node[hidden neuron] (H_4-\name) at (5*\layersep,{-\y-div(\y,12)}) {};
			
	\begin{scope}[local bounding box=softmax1]
		\foreach \name / \y in {1,...,8}
		{	\pgfmathparse{\y<8 ? \y : "n"}
			\path[yshift=5.5cm]node[output neuron,pin={[pin edge={->}]right:\rotatebox{270}{$y_{\pgfmathresult}$}}] (O-\name) at (6.5*\layersep,{-\y-div(\y,8)}) {};
		}
	\end{scope}
	
	\begin{scope}[local bounding box=softmax2]
		\foreach \name / \y in {9,...,16}
		{	\pgfmathparse{\y<16 ? int(\y-8) : "n"}
			\path[yshift=6.5cm]node[output neuron,pin={[pin edge={->}]right:\rotatebox{270}{$y_{\pgfmathresult}$}}] (O_2-\name) at (6.5*\layersep,{-\y-div(\y,16)-6}) {};
		}
	\end{scope}

		\foreach \source in {1,...,8}
        \foreach \dest in {1,...,12}
            \path (I-\source) edge (H_1-\dest);
		\foreach \source in {1,...,12}
        \foreach \dest in {1,...,12}
            \path (H_1-\source) edge (H_2-\dest);
		\foreach \source in {1,...,12}
        \foreach \dest in {1,...,12}
            \path (H_2-\source) edge (H_3-\dest);
		\foreach \source in {1,...,12}
        \foreach \dest in {1,...,12}
            \path (H_3-\source) edge (H_4-\dest);
		\foreach \source in {1,...,12}
			\foreach \dest in {1,...,8}
				\path (H_4-\source) edge (O-\dest);
		\foreach \source in {1,...,12}
			\foreach \dest in {9,...,16}
				\path (H_4-\source) edge[line width=0.1pt] (O_2-\dest);

  	\node[annot,rotate=-90,below of=I-4, node distance=2cm] (il2) {Input layer};
    \node[annot,rotate=-90,above of=O-5] {Output layer};
		\node[annot,rotate=-90,above of=O_2-13] {Output layer};
		\draw[surround] ($(softmax1.north west) + (-15.5pt, 15.5pt)$) rectangle
        ($(softmax1.south east) + (2.5pt, -15.5pt)$);
		\draw[surround] ($(softmax2.north west) + (-15.5pt, 15.5pt)$) rectangle
        ($(softmax2.south east) + (2.5pt, -15.5pt)$);
	
\end{tikzpicture}
\end{turn}
\captionsetup{justification=centering}
\caption{Pronunciation Variability Modeling \\ \small{Neuron color scheme: Red-Output, Blue-Hidden, Gray-ivector input, Green-MFCC input}}
\label{fig:pv}
\end{figure*}

\subsection{Accounting for Acoustic Variability}
We assume that acoustic variability affects the lower-level network structures only and hence these layers need to be adapted to better represent the children's feature subspace.
This could be thought of as retaining the knowledge of higher level abstract functions (mappings) from an adult's ASR, while accounting for the spectral variabilities.
This parallels alternate approaches such as feature space transforms like VTLN, fMLLR.
One important difference is the degrees of freedom and hence parameters that this technique allows, likely resulting in better transformations but also much larger demands on adaptation data.
Hence, to account for the acoustic variability we retain all the hidden layers from adult models except the bottom-most layer as shown in Figure~\ref{fig:av}.
The DNN is retrained with children speech until convergence to estimate the optimal parameters of the lowest layer.
This find is interesting as most of the transfer learning techniques adapt the output layers \cite{tong2017transfer, tong2017multi,serizel2014deep,serizel2017deep} while for this task we adapt the input layer(s).

Moreover, we also augment the MFCC features with i-vector information. 
The i-vector subspace has been shown to capture speaker specific information efficiently \cite{dehak2011front}.
It has also been successfully used for capturing speaker age characteristics \cite{shivakumar2014simplified}.
Further speaker specific information is useful for speaker adaptation of DNN acoustic models \cite{saon2013speaker}.
The augmentation of i-vectors enables for better adaptation of the bottom layers during transfer learning by estimating speaker and age specific spectral transformations which are highly relevant for modeling children speech.

\subsection{Accounting for Pronunciation Variability}

We assume that phonemic variability affects the higher-level network structures only and hence these layers need to be adapted to better represent the children's pronunciation variance.
Hence we propose to adapt higher layers towards modeling pronunciations as illustrated in Figure~\ref{fig:pv}.
This parallels work in adapting acoustic models across languages \cite{heigold2013multilingual, huang2013cross} or for non-native speakers \cite{matassoni2018}.
In this case we are only tackling the pronunciation variability and as such the lower-order layers will remain unchanged.

\subsection{Accounting for Acoustic \& Pronunciation Variability}
\label{sec:disjoint_training}
Finally, to account for both the acoustic and pronunciation variability, we would like to update both the top-most and bottom-most layers and keep the rest of the layers fixed.
This is attempted in two ways: (i) keeping weights of the middle hidden layers fixed and allow the top-most and bottom-most layer(s) to update simultaneously,
(ii) dis-jointly and alternately training the various layers (top \& bottom) until convergence.
The motivation behind the disjoint training is to constrain the updatable parameters at any time, to limit the adaptation, and to regulate the amount of knowledge retained from adult acoustic models.

\begin{table*}[!b]
\renewcommand{\arraystretch}{1.3}
\renewcommand\tabularxcolumn[1]{>{\Centering}p{#1}}
\newcolumntype{b}{X}
\newcolumntype{s}{>{\hsize=.5\hsize}X}
\centering
\begin{tabularx}{0.75\textwidth}{| b | s | s | s | s |}
\hline
Corpus & \# Hours & \# Speakers & Age & Split\\
\hline
CU Prompted \& Read & 25.69 & 663 & 6-11 & Train \\
CU Read \& Summarized & 33.11 & 320 & 6-11 & Train \\
OGI & 22.56 & 509 & 6-11 & Train \\
ChIMP & 10.25 & 97 & 6-14 & Train \\
\hline
CID & 2.26 & 324 & 6-14 & Test \\
\hline
\hline
Total (Children-Train) & 91.61 &  1589 &  6 - 14 & Train \\
\hline
\hline
\hline
TED-LIUM (Adult) & 205.82 & 774 & NA & Train \\
\hline
\end{tabularx}
\caption{Summary of Corpora and their training-testing splits} \label{tab:split}
\end{table*}

\section{Databases}
\label{sec:database}
In this work we employ 5 different children speech databases and 1 adult speech corpora. All the data are processed at 16kHz.

\begin{figure*}[t]
\centering
\includegraphics[width=0.75\textwidth,height=0.5\textwidth]{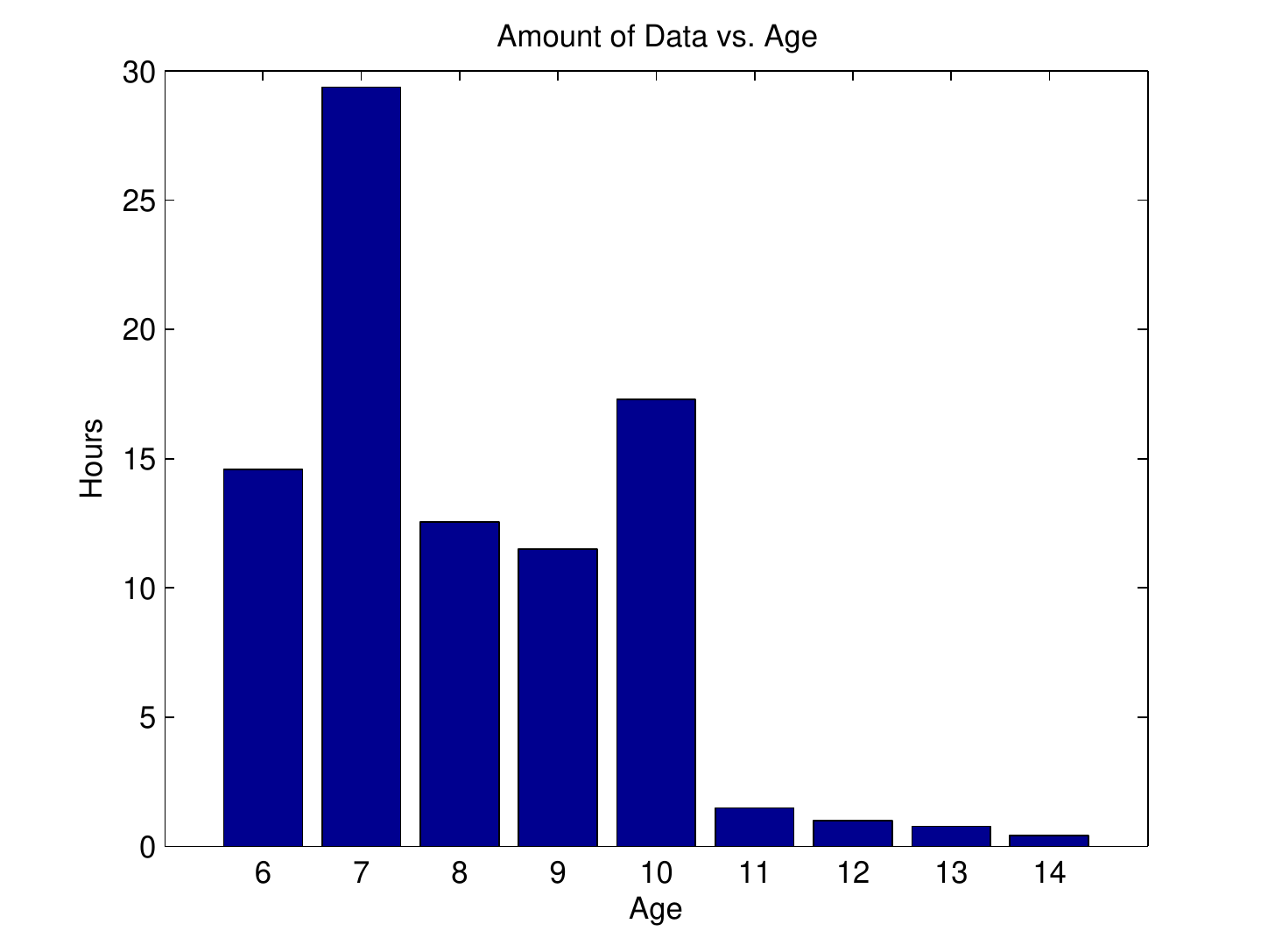}
\caption{Distribution of training data over Children Age}\label{fig:age_duration_distribution}
\end{figure*}

The following children speech databases were used:
\begin{enumerate}
\item CU Kid's Prompted and Read Speech Corpus \cite{cole2006university2}
\item CU Kid's Read and Summarized Story Corpus \cite{cole2006university}
\item OGI Kid's Speech Corpus \cite{shobaki2000ogi}
\item ChIMP Corpus \cite{potamianos1998spoken}
\item CID Children's Speech Corpus \cite{lee1999acoustics}
\end{enumerate}
Using multiple children's speech corpora makes the problem more challenging and more relevant to real world scenarios.
The CID Children's Speech Corpus is used for testing and the rest for training.
The summary of breakup of databases and their split for training and testing is provided in table ~\ref{tab:split}. The distribution of data over the age is illustrated in Figure~\ref{fig:age_duration_distribution}.

The adults corpus employed in this work is the TED-LIUM ASR corpus \cite{rousseau2014enhancing}.
It consists a total of 206 hours of speech data of 774 speakers giving TED talks.

\section{Experimental Setup \& Baseline System}
\label{sec:baseline_setup}

\subsection{Experimental Setup}
The experimental setup is very similar to the one used in our previous work \cite{shivakumar2014improving}.

\subheading{GMM-HMM System:}
We employ as a baseline a Gaussian Mixture Model based Hidden Markov Model ASR.
For this system the features used are standard Mel-Frequency Cepstral Coefficients (MFCC) of dimension 13 with window size of 25ms and shift of 10 ms with their first order and second order derivatives.
The HMMs were modeled using 3 states for non-silence phones and 5 for silence phones.
We also employ the front-end adaptation techniques of Linear Discriminant Analysis (LDA), Maximum Likelihood Linear Regression (MLLR) and Feature space MLLR (fMLLR) for speaker independent and speaker adaptive training.

\subheading{Dictionary:}
We employ the CMU Pronunciation dictionary \cite{weide1998cmu}.
This dictionary corresponds to American-English pronunciations and that makes it compatible with our available children and adult data.
To account for the out-of-vocabulary (OOV) words during training, a grapheme to phoneme converter was used to generate phoneme transcripts for OOV words. 

\subheading{Language Model:}
Two language models were interpolated, one trained on a subset of children's training data reference transcripts 
and the second generic English language model from CMU-Sphinx-4\footnote{Language model version: cmusphinx-5.0-en-us.lm} \cite{walker2004sphinx}. 
The interpolation helps incorporating children's grammar which is beneficial for children's ASR along with the adult's grammar to facilitate the transfer learning process between adults and children.
Since this work deals with evaluating acoustic models, we keep the language model fixed for all our experiments. 

\subheading{i-Vector Setup:}
We employ high-resolution, 40-dimensional MFCCs as front-end features for i-vector training.
To introduce context, we used an LDA transform with a context of 3 left and 3 right.
Both the universal background model (UBM) and the total-variability matrix for the i-vector were trained on adults speech data to allow transfer learning from these as well.
We used 2048 Gaussian components to train the UBM, whereas the i-vector dimension was fixed to 100.

\subheading{Hybrid DNN-HMM System:}
We employed a hybrid DNN-HMM system, where the DNN is used to replace the posterior probabilities of a traditional GMM system.
DNN architecture employed is a time delay neural network which uses sub-sampling for exploiting long contextual information \cite{peddinti2015time}.
The DNN consumes high resolution MFCC features with a context of 13 left and 9 right frames.
The MFCC features were concatenated with the i-vector and were used to train the DNN.
The DNN has 7 hidden layers, each of dimension 3500.
p-norm non-linearity was used in the hidden layers.
The output Softmax layer consists of 3976 units trained to predict the posterior.
We used greedy layer-wise training to train the DNN \cite{bengio2007greedy}. 

\subsection{Baseline System}
\subsubsection{Children's ASR}
The Children's ASR was trained only on the children speech data (splits illustrated in table~\ref{tab:split}).
In order to compare to the DNN and to relate to the previous work \cite{shivakumar2014improving}, we provide the result of the GMM-HMM systems. 
To asses the advantage of the proposed transfer learning, we also trained a hybrid DNN-HMM based baseline system on children-only speech data.
To provide a range of baselines we also employ popular adaptation techniques such as VTLN, SAT, i-vector,  which have been proven successful for children's speech, in conjunction with the Hybrid DNN-HMM.

\subsubsection{Adult's ASR}
An additional ASR was trained only on adults speech data from TED-LIUM.
The performance of this system was evaluated by decoding on the test set of children speech to compare its performance to that of the baseline children ASR.
This system is used for transfer learning to adapt to children speech.

\begin{table*}[!t]
\renewcommand{\arraystretch}{1.3}
\renewcommand\tabularxcolumn[1]{>{\Centering}p{#1}}
\newcolumntype{b}{>{\hsize=1.5\hsize}X}
\newcolumntype{s}{>{\hsize=.5\hsize}X}
\centering
\begin{tabularx}{0.75\textwidth}{| b | s |}
\hline
\textbf{Model} & \textbf{WER} \\
\hline
GMM-HMM Monophone & 54.53\% \\
GMM-HMM Triphone & 36.96\% \\
GMM-HMM Triphone LDA+MLLT & 32.79\% \\
GMM-HMM Triphone LDA+MLLT+SAT & \textbf{24.55\%} \\
GMM-HMM Triphone LDA+MLLT+SAT + VTLN & 25.66\% \\
\hline
Hybrid DNN-HMM & 35.97\% \\
Hybrid DNN-HMM + VTLN & 32.72\% \\
Hybrid DNN-HMM + LDA+MLLT+SAT & \textbf{21.31\%} \\
Hybrid DNN-HMM + LDA+MLLT+SAT + VTLN & 21.82\% \\
Hybrid DNN-HMM + online i-vector (speaker) & 28.03\% \\
Hybrid DNN-HMM + online i-vector (utterance) & 26.59\% \\
Hybrid DNN-HMM + offline i-vector (utterance) & 25.53\% \\
\hline
\end{tabularx}
\caption{Baseline results of ASR trained only on children's speech (91 hours).}\label{results:baseline}
\end{table*}

\section{Recognition Results and Discussions}\label{sec:results}
\subsection{Baseline Results}
Table~\ref{results:baseline} shows the results of the baseline system.
The GMM-HMM results are comparable to that of the previous study \cite{shivakumar2014improving} although more data has been incorporated for training in the current system.
We see that the SAT gives the best results among the GMM-HMM framework.
The hybrid DNN-HMM system improves over its respective GMM counterpart by 1\% absolute.
We believe the reason for the minimal improvement is that DNN requires more data to generalize well for children speech.

We also compare different adaptation techniques for the DNN-HMM model.
VTLN provides an absolute 3.25\% improvement over the raw MFCC features.
SAT performs much better and reduces the WER to 21.31\% an absolute improvement of 14.66\% over raw features.
However, we find that a combination of VTLN and SAT doesn't provide any major improvement.
Trials augmenting raw features with i-vectors suggest that the best performance is achieved by using the offline version of i-vectors calculated on the whole utterance.
However, these still fail to surpass the performance of the SAT by 4.22\% absolute, thereby confirming SAT is crucial for children speech adaptation irrespective of GMM or DNN acoustic modeling.

\begin{table*}[!b]
\renewcommand{\arraystretch}{1.3}
\renewcommand\tabularxcolumn[1]{>{\Centering}p{#1}}
\newcolumntype{b}{X}
\newcolumntype{x}{>{\hsize=.25\hsize}X}
\newcolumntype{s}{>{\hsize=.5\hsize}X}
\centering
\begin{tabularx}{0.75\textwidth}{| b | x | x | b | s |}
\hline
\textbf{Model} & \textbf{AV} & \textbf{PV} & \textbf{Configuration} & \textbf{WER} \\
\hline
DNN Children & \xmark & \xmark & Baseline & 25.53\% \\
DNN Adult & \xmark & \xmark & Baseline &  39.32\% \\
DNN Children + Adult & \xmark & \xmark & - & 20.35\% \\
DNN TL & \xmark & \cmark & 1 layer & 26.97\% \\
DNN TL & \cmark & \xmark & 1 layer & 24.26\% \\
DNN TL & \cmark & \cmark & 1 layer each & \textbf{19.63\%} \\
DNN TL & \cmark & \cmark & dis-joint 1 layer each & 20.01\% \\
DNN TL & \cmark & \cmark & 2 layers each & \textbf{17.8\%} \\
DNN TL & \cmark & \cmark & dis-joint 2 layers each & 18.74\% \\
DNN TL & - & - & all layers & \textbf{17.8\%} \\
\hline
\end{tabularx}
\captionsetup{justification=centering}
\caption{Transfer Learning Results (DNN: Hybrid DNN-HMM + offline i-vector (utterance level)\\ AV: Acoustic Variability Modeling, PV: Pronunciation Variability Modeling)}\label{results:main}
\end{table*}

\subsection{Transfer Learning Results}
\label{sec:results_tl}
\Pg{Single layer adaptation}

Table~\ref{results:main} shows  results of the proposed transfer learning technique.
The baseline adult's model is significantly worse than children's model, as expected and consistent with previous studies.

We first conduct adaptation experiments by adapting a single layer at a time.
This allows us to assess the types of variability present in children's speech relative to the adult-trained DNN.
It also allows us to evaluate performance benefits through addressing specific variability types.
Adapting bottom layers should help counter acoustic variability in kids.
Adapting top layers should attempt to account for pronunciation variability.

We observe as hypothesized that with single-layer modifications addressing acoustic variability (24.26\%) is more advantageous than accounting for pronunciation variability (26.97\%).
Both are providing big gains over both the original adult's baseline of 39.32\%.

Often, in transfer learning the top layers, representing high level abstract information, are used for adaptation \cite{huang2013cross,heigold2013multilingual}. 
However, our finding is in agreement with prior studies showing high variability in spectral characteristics of children speech \cite{potamianos2003robust,lee1999acoustics,potamianos1997automatic,li2002analysis} that denotes the need for input-layer adaptation.
This suggests that the transfer learning adaptation configuration is task dependent.


\section{Analysis of Amount of Adaptation Data}\label{sec:adaptation_data}
\begin{figure*}[t]
\centering
\begin{tikzpicture}
   
		\node[anchor=south west,inner sep=0] at (0,0) {
		\includegraphics[width=0.75\textwidth,trim={0cm 0cm 0cm 0cm},clip]{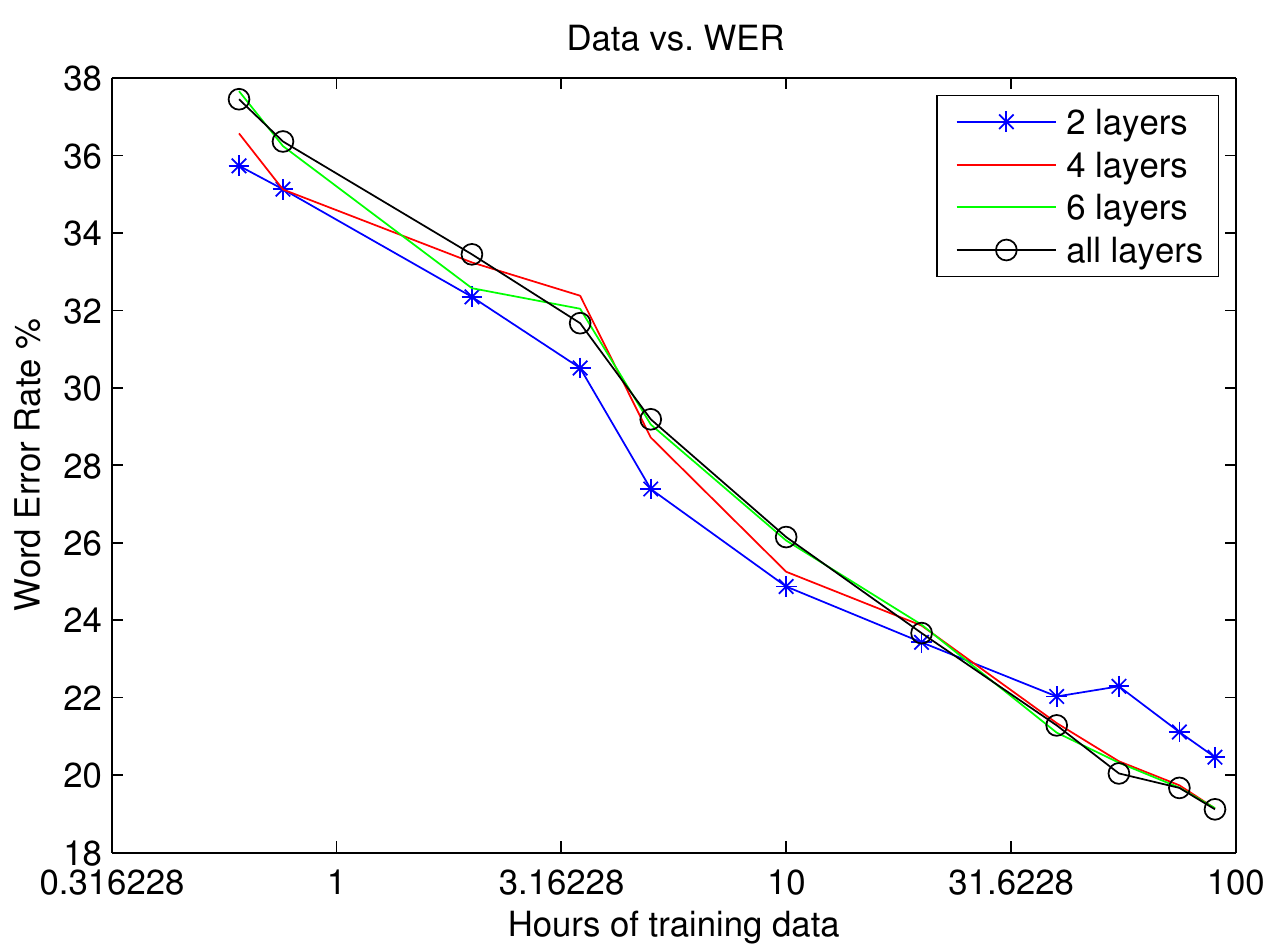}
		};
    \draw[thick,dotted] (10.35,1.05) -- (10.35,3.2);
\end{tikzpicture}
\caption{Amount of Adaptation Data (Log-scale) versus Word Error Rate; Four Different DNN configurations}
\label{fig:graph}
\end{figure*}

\begin{table*}[!b]
\renewcommand{\arraystretch}{1.3}
\renewcommand\tabularxcolumn[1]{>{\Centering}p{#1}}
\newcolumntype{b}{>{\hsize=.75\hsize}X}
\newcolumntype{s}{>{\hsize=.5\hsize}X}
\centering
\begin{tabularx}{0.75\textwidth}{| b | b | s |}
\hline
\textbf{Adaptation Data} & \textbf{Model (training)} & \textbf{WER} \\
\hline
35 minutes & 2 layers (simultaneous) & 35.73\% \\
35 minutes & 2 layers (dis-joint) & 35.04\% \\
45 minutes & 2 layers (simultaneous) & 35.13\% \\
45 minutes & 2 layers (dis-joint) & 34.33\% \\
2 hours & 2 layers (simultaneous) & 32.35\% \\
2 hours & 2 layers (dis-joint) & 32.94\% \\
\hline
\end{tabularx}
\caption{Adaptation at extreme low data scenarios}\label{tab:low_data}
\end{table*}

\Pg{Multiple layer adaptation}
We also investigate  letting both the top and bottom layers update, i.e., by modeling both the acoustic and pronunciation variability simultaneously.
We observe a further boost in accuracy with the WER dropping to 19.63\% giving a relative gain of 23.1\% over the baseline children model and 50.1\% over adults model.
One interesting observation is that the performance benefits achieved by simultaneously updating the top and bottom layer is complementary to that achieved by adapting each layers individually.
This suggests that \emph{the acoustic variability and the pronunciation variability are fairly exclusive of each other in case of children}.
Dis-joint training doesn't provide improvements, likely due to the sufficient amounts of data to simultaneously account for the degrees of freedom of joint training.
It could however be beneficial in the case of less data as we show in the subsequent section (Section~\ref{sec:low_data}).

In our experiments we also found that using 2 layers to update instead of 1 gives further improvements.
We achieve a word error rate (WER) of 17.8\% which is a modest 9.3\% gain over using single layers for adaptation.
Subsequent experiments with more layers did not provide any significant improvements.
Adapting all the layers gives the same performance of 17.8\% WER.
This suggests that all the variability present between the children and the adult is concentrated at the top (pronunciation level) and bottom (acoustic level) layers of the DNN in agreement with the initial hypothesis made in this work.
This indicates that the underlying middle hidden layers efficiently model the basic human speech structure.

Overall, the proposed transfer learning technique outperforms the best results obtained using the baseline model trained on children's speech with SAT by a relative 16.5\% (relative 54.7\% improvements over the baseline adult model).
The results highlight the power of transfer learning in the DNN framework in outperforming SAT, the prior best performing recipe for children ASR \cite{shivakumar2014improving}.


Finally, we compare the proposed adaptation technique against a model trained on combined data of adult's and children's speech which was proposed in \cite{qian2016mismatched,liao2015large,fainberg2016improving,qian2016improving}.
Combining adults' and children's data provides modest improvements over the baseline systems trained only on children (5.18\% absolute) and adult (18.97\% absolute) data.
However, our proposed adaptation technique proves to be superior with 2.55\% absolute improvement over the model trained on adults and children.

Informed by the above results, for the rest of this work, we experiment with four different adaptation configurations:
\begin{enumerate}
\item 2 layers: (bottom-most + top-most) 
\item 4 layers: (2-bottom-most + 2-top-most)
\item 6 layers: (3-bottom-most + 3-top-most)
\item all layers.
\end{enumerate} 
We always adapt even number of layers, thus maintaining symmetry in the structure in terms of top and bottom layers for maximum performance.
Moreover, from our experiments we found that adapting a single layer never surpasses the adaptation using symmetric 2 layers and thus we skip presenting those results.

\begin{figure*}[!t]
\centering
\includegraphics[width=0.85\textwidth, height=0.5\textwidth]
{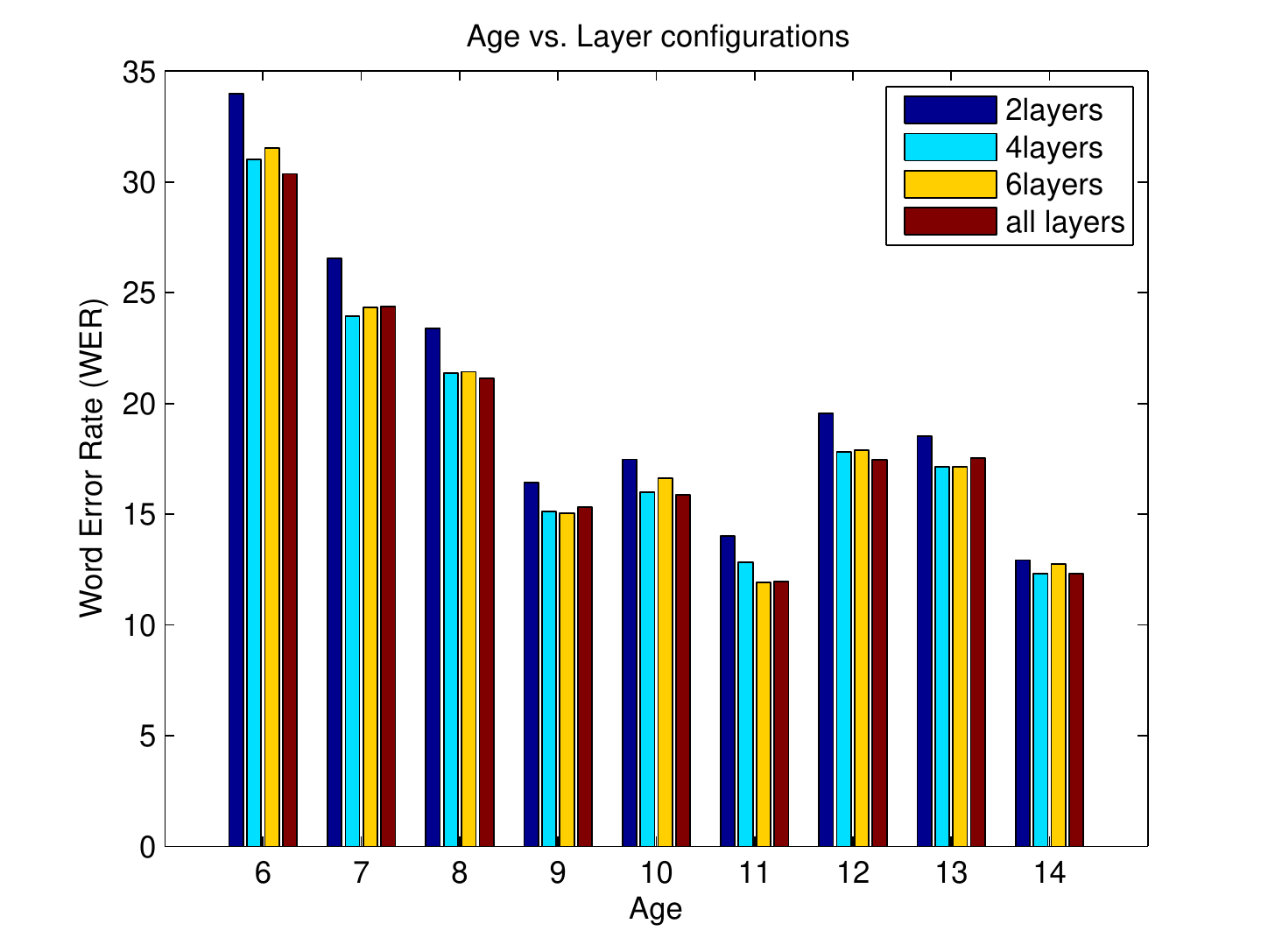}
\caption{Children Age vs. Adaptation layer configurations}\label{fig:adapt_config_vs_age}
\end{figure*}


Figure~\ref{fig:graph} shows the transfer learning adaptation performance curve over amount of adaptation data (in terms of WER and hours).
Each curve represents different adaptation architectures of the DNN in terms of number of layers used for adaptation. The following inferences can be drawn from the plot:
\begin{itemize}

\item The WER decays \emph{exponentially} with increase in amount of data.

\item The curves are almost always monotonically decreasing, suggesting that more adaptation data always helps.
  We note that the graph has not converged, meaning more data could help the adaptation further, suggesting that the constraint is still the amount of children data available.

\item Any amount of children data is helpful for adaptation, as in our experiments even as low as 35 minutes of children adaptation data was found to give improvements of up-to 9.1\% (relative) over the adult model.
  
\item Adapting less number of layers yields better results for low data scenario, i.e., we find that adapting only 2 layers consistently outperforms adapting with more layers until about 25 hours of adaptation data.
  
\item With 25 hours of adaptation data, all of the 4 curves more or less intersect suggesting that all the four architectures gives approximately the same improvements.
  
\item For more than 25 hours of data, we find that adapting 4, 6 and all layers converge to approximately same performance in agreement of the findings in section~\ref{sec:results_tl}.
  
\end{itemize}


\subsection{Transfer Learning for low resource scenarios}\label{sec:low_data}

Table~\ref{tab:low_data} represents three extreme low data adaptation scenarios.
We apply dis-joint training to account for data sparsity as explained in section~\ref{sec:disjoint_training}.
Since earlier experiments indicated that 2 layers provided maximum benefits for low data, we present the effect of dis-joint training for 2 layers only.
The series of experiments involved first training with top and bottom layer and fixing those weights.
We then continue training with layer 2 and 6 to update.
We find that the dis-joint training further improves the adaptation for small amounts of data i.e., 35 and 45 minutes.
The improvements diminish when more data is used, as in the case of 2 hours and as seen earlier in table~\ref{results:main}.
Approximately 1.9\%  and 2.3\% relative reduction in WER is observed for 35 and 45 minutes respectively.

\section{Age dependent analysis}\label{sec:age_analysis}

\subsection{Age vs. Adaptation layer configurations}\label{sec:layerconfig_age}

\begin{figure}[!b]
\centering
\includegraphics[width=1.0\columnwidth,trim={0 0 0 2cm}]
{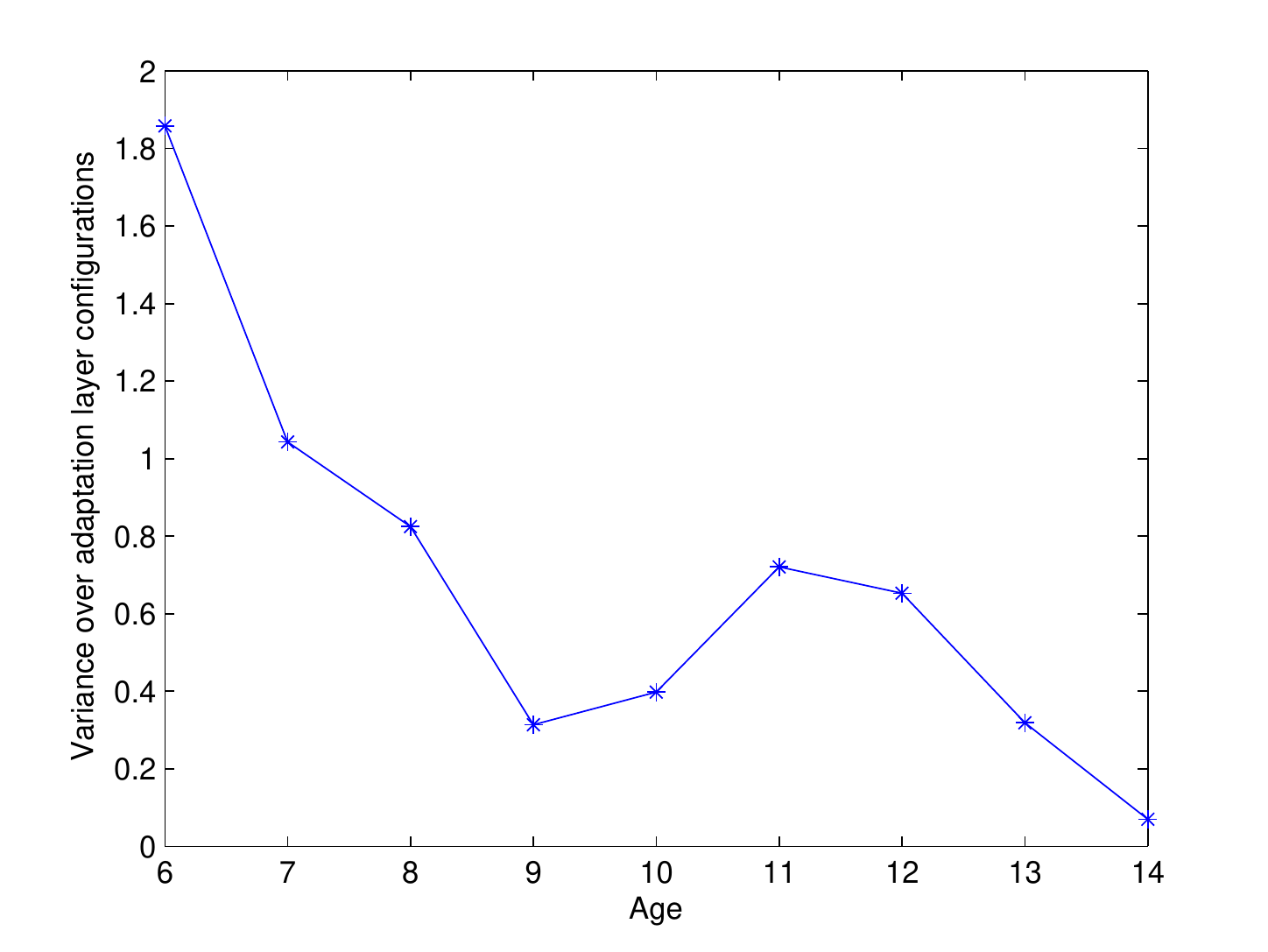}
\captionsetup{justification=centering}
\caption{WER Variance over Adaptation Layer Configurations \\across Children Age Groups}\label{fig:adapt_config_variance}
\end{figure}

In this section, we analyze the effect of different layer adaptation configurations on the children's age.
The model is trained on all available children data independent of age (age-independent acoustic model).
The results are plotted as a bar graph in Figure~\ref{fig:adapt_config_vs_age}.
We observe the following:
\begin{itemize}

\item Overall performance increases with increase in age, irrespective of the adaptation configuration.
  The two peaks corresponding to ages 12 and 13 years is probably a consequence of the acoustic model mismatch posed by relatively less training data for elder children (11 - 14 years) (See Figure~\ref{fig:age_duration_distribution}).
  
\item Performance is worse for younger children,  consistent with past work \cite{shivakumar2014improving}.
  
\item The adaptation configuration affects more  younger children.
  To demonstrate this, Figure~\ref{fig:adapt_config_variance} shows the WER variance between the 4 configurations plotted over age. It is evident from the plot that the variance for younger children is significantly higher and decreases with increase in age.
  Similar peaks found in Figure~\ref{fig:adapt_config_vs_age} for ages 12 and 13 years is also apparent in variance plot.
  
  \item Younger children benefit with adaptation of more layers than older children.
  This aligns with the expectation that younger children manifest higher acoustic complexity and hence more parameters (layers) are necessary to capture the increased complexity.
  For example, from Figure~\ref{fig:adapt_config_vs_age}, if 2 layers are adapted rather than all layers we have significantly fewer gains for 6 year olds than 14 year olds.
  This is also justified to certain extent by looking at the variances in Figure~\ref{fig:adapt_config_variance}.
  This also suggests that despite the acoustic and pronunciation variability, young-children speech encodes more variability that affects the whole network.
\end{itemize}

\begin{figure*}[!t]
\centering
\includegraphics[width=0.75\textwidth,height=0.44\textwidth,trim={2.5cm 1cm 3cm 2.1cm},clip]{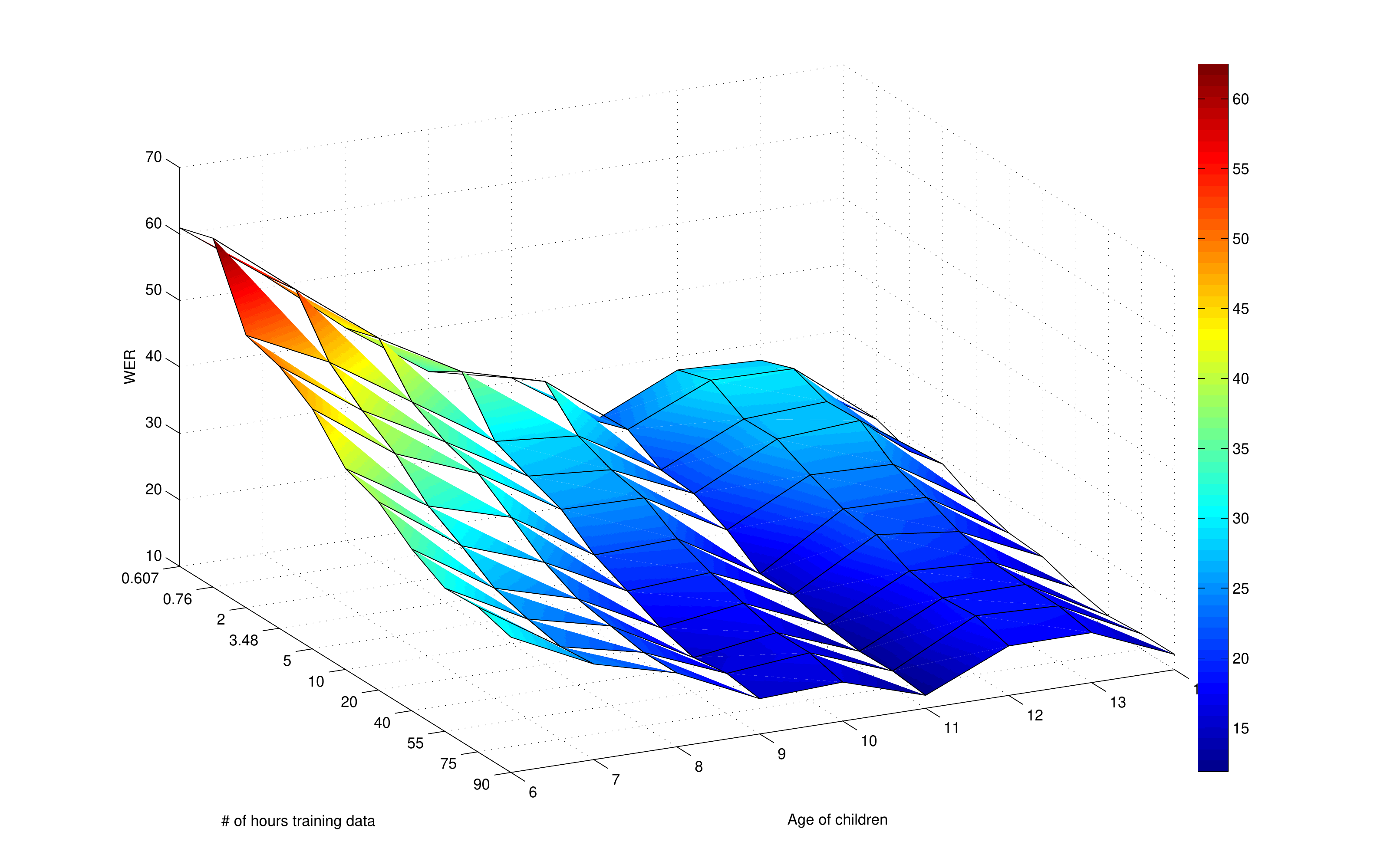}
\caption{Amount of training data vs. Children Age}\label{fig:data_age}
\end{figure*}

\subsection{Amount of Adaptation Data vs. Age}\label{sec:data_age}

We also investigate the amount of adaptation data and its effect on children's age.
Figure~\ref{fig:data_age} shows a 3-d plot of WER over the amount of adaptation data and the children's age.
Adaptation data are chosen at random and hence follow the proportions in Figure~\ref{fig:age_duration_distribution}.
We make the following inferences from the figure:
\begin{itemize}
\item It is evident that more the adaptation data better is the performance irrespective of age of the children.

\item We see that younger children need more data to reach the same level of performance as older children.
  The trend is in accordance with the age, i.e., as the age of children increases, less amount of adaptation data is sufficient.
  
\item In-spite of large amount of matched-adaptation data, we observe that the performance of younger children of age 6-8 years doesn't meet that of the elder children.
  
\item Although the adaptation data for older children is mainly mismatched (see Figure~\ref{fig:age_duration_distribution} for distribution of training data), they need as low as 30 minutes of adaptation data to surpass the performance of the younger children adapted on all (90 hours) of data.
\end{itemize}

\subsection{Layer configurations vs. Amount of Adaptation Data vs. Age}\label{sec:layerconfig_data_age}
\begin{figure*}[!t]
\centering
\includegraphics[width=0.75\textwidth,height=0.5\textwidth,trim={2.5cm 0cm 3cm 1cm},clip]{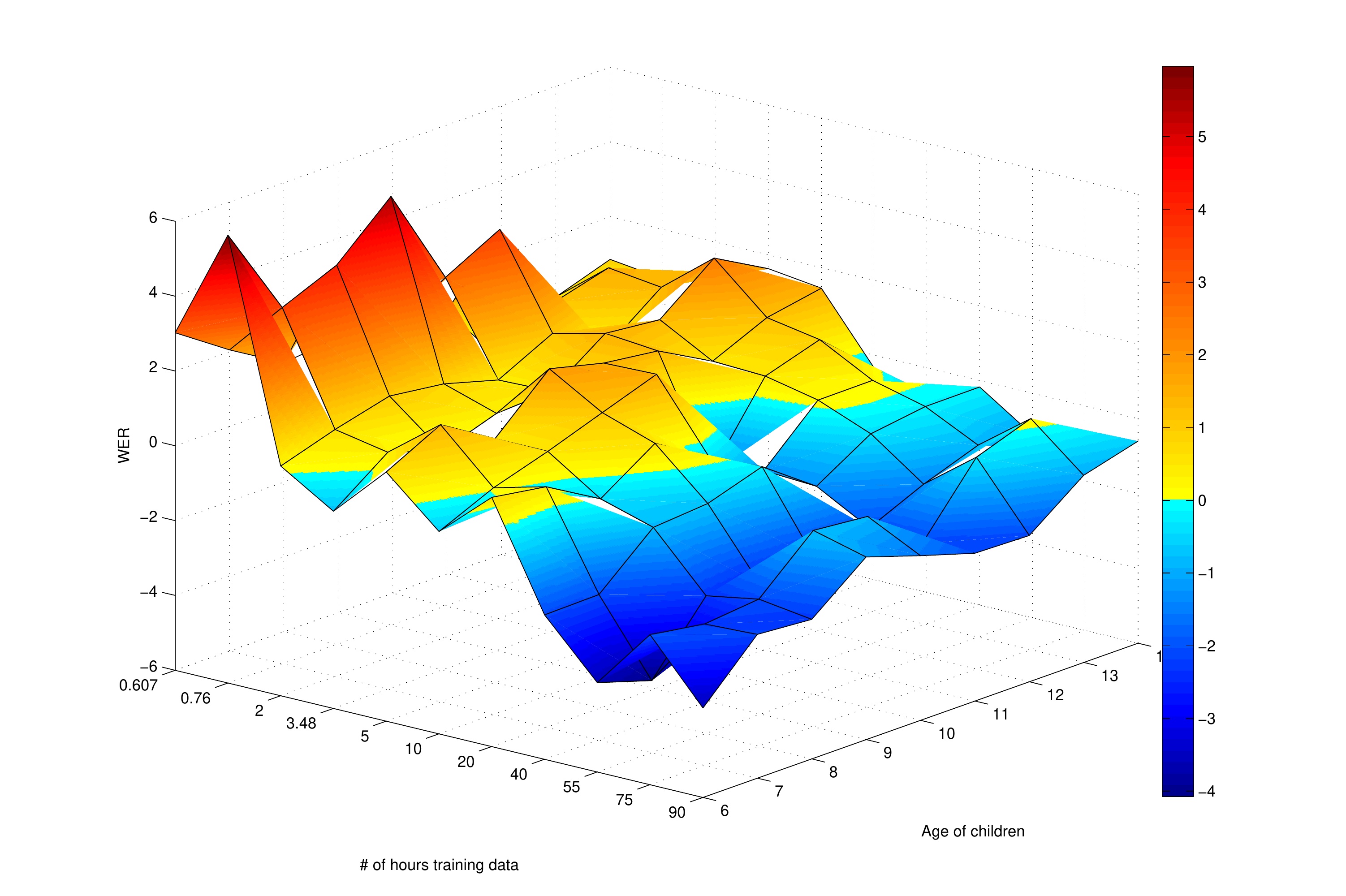}
\caption{Layer configurations vs. Amount of Adaptation Data vs. Age}\label{fig:data_age_layerconfig}
\end{figure*}
To gain insights into the optimal adaptation strategy in terms of 4 earlier mentioned adaptation layer configurations as a function of the amount adaptation data and age of children, we plot the difference of WER between different adaptation layer configurations.
Figure~\ref{fig:data_age_layerconfig} shows a 3-d plot for difference between the WER when adapting all layers and WER when adapting only 2 layers.
Any positive values indicate that adapting with 2 layers to be superior than adapting all the layers and vice-versa.
We can deduce the following by looking at Figure~\ref{fig:data_age_layerconfig}:
\begin{enumerate}
\item Adapting 2-layers is more beneficial when adaptation data available is low.
  When more adaptation data is available, it is advantageous to adapt more layers.
  The trend is consistent over all the children ages - 6 years to 14 years which is in accordance with the finding from Section~\ref{sec:layerconfig_age} and Section~\ref{sec:data_age}.
  
\item For younger children, 6 years to 11 years, we find that it is better to use fewer adaptation layers when the adaptation data available is low.
  The performance of the system is significantly lower when adapting with all the layers.
  This is because of the increased variability affecting the overall performance of the system.
  This is especially true when a large amount of parameters are adapted with little data, due to noise introduced from high variability.
  The performance of the system eventually recovers and surpasses the 2-layer adaptation configuration when sufficient amount of adaptation data is available.
  
\item For younger children, with sufficiently high adaptation data, we find that the effective gains made between the layer configuration is much higher compared to elder children.
  Thereby asserting their sensitivity to adaptation data and layer configurations.
  
\item For older children, 12 years to 14 years, the system adapts rapidly with considerably less data compared to younger children.
  
\item For older children, the performance gains are comparable between  2-layer adaptation and all layer adaptation.
\end{enumerate}

The analysis is only presented for differences between adapting all the layers and adapting only 2-layers.
The particular plot was chosen to illustrate the differences as in an extreme case.
Similar trends were observed for differences of other configurations, i.e., adapting more layers versus fewer layers. 

\section{Analysis of Age Dependent Transformations}\label{sec:age_dependent}

\begin{figure*}[t]
\centering
\includegraphics[width=0.75\textwidth,height=0.5\textwidth,trim={2.5cm 0cm 3cm 1cm},clip]{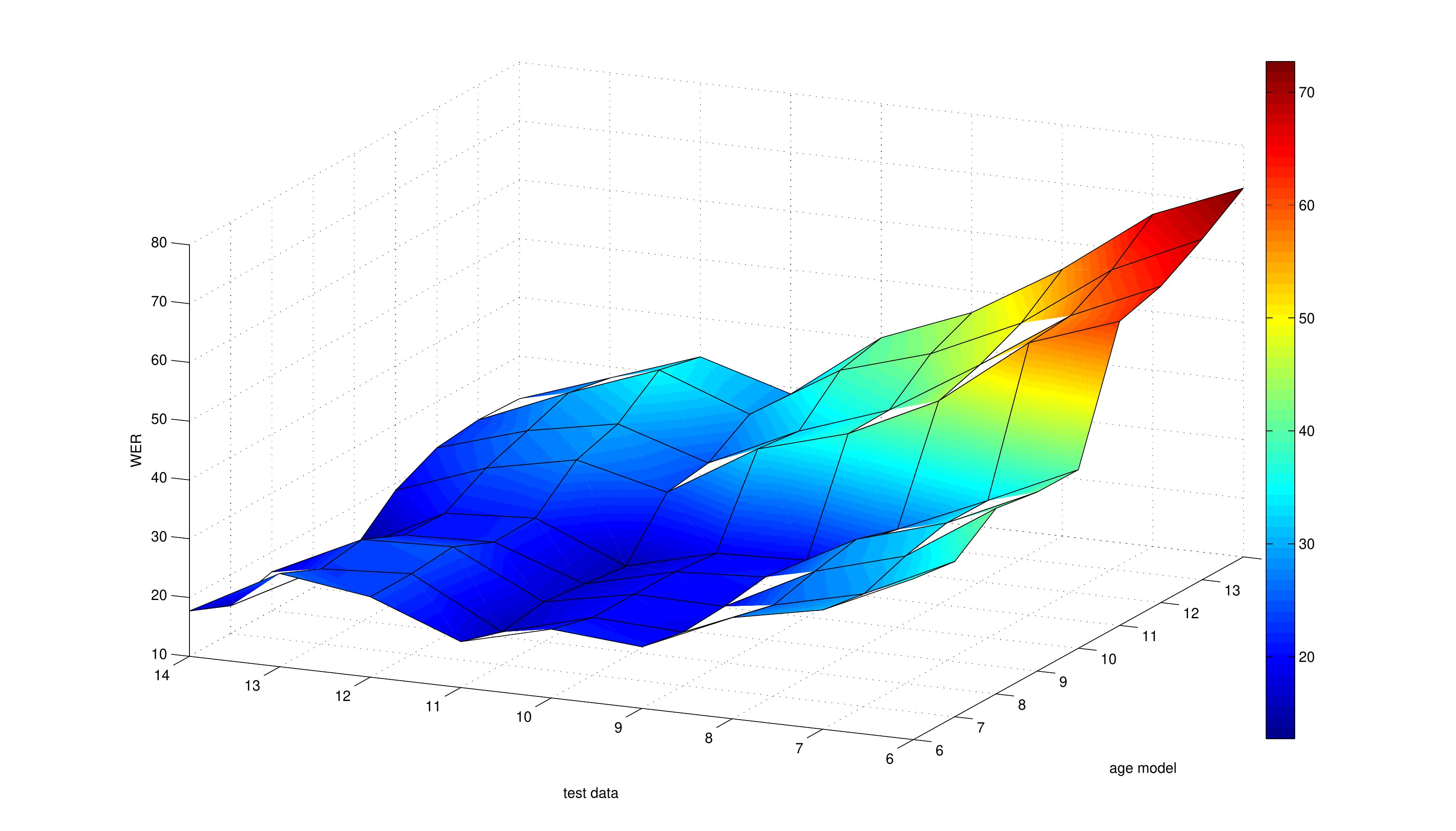}
\caption{Age dependent model performance - Adapting all layers}\label{fig:age_dep_all_layers}
\end{figure*}

In order to assess the validity of the transformations learnt by adapting the layers and its extensibility and relevance to children's speech, we analyze age specific transformations resulting from age dependent transfer learning adaptation. The transformations would be meaningful if there exists some level of meaningful portability between different ages. Note: these transformations are not equivalent and shouldn't be mistaken to age specific models. 

Figure~\ref{fig:age_dep_all_layers} shows the 3-d plot of WER from application of age dependent transformations on each age group, when adapting the model with all the layers. The following can be inferred from the plot:
\begin{enumerate}
\item For younger children, ages 6 years to 10 years, the matched models i.e., application of same aged transformations provide significant improvements.
\item For younger children, as the mismatch increases (in terms of age), the performance decreases.
\item For younger children, the rate of performance degradation is much more drastic as the mismatch (in terms of age) increases compared to older children.
\item For ages 11 years to 14 years, the surface is more or less plateaued, this is probably because of data scarcity for estimation of meaningful transformations (See Figure~\ref{fig:age_duration_distribution}).
\item The overall surface is tilted towards the left, indicating that performance of elder children are significantly better irrespective of the applied transformation.
\end{enumerate}

The above observations confirm the validity of the transformations and its portability across the ages. Although the transformations are not equivalent to age-dependent models, the above observations prove they exhibit similar trends (performance-wise) as reported in previous literature \cite{elenius2005adaptation}.

\subsection{Age dependent transformations versus Adaptation layer configurations}

\begin{figure*}[t]
\centering
\includegraphics[width=0.75\textwidth, trim={2.5cm 0cm 3cm 1cm},clip]{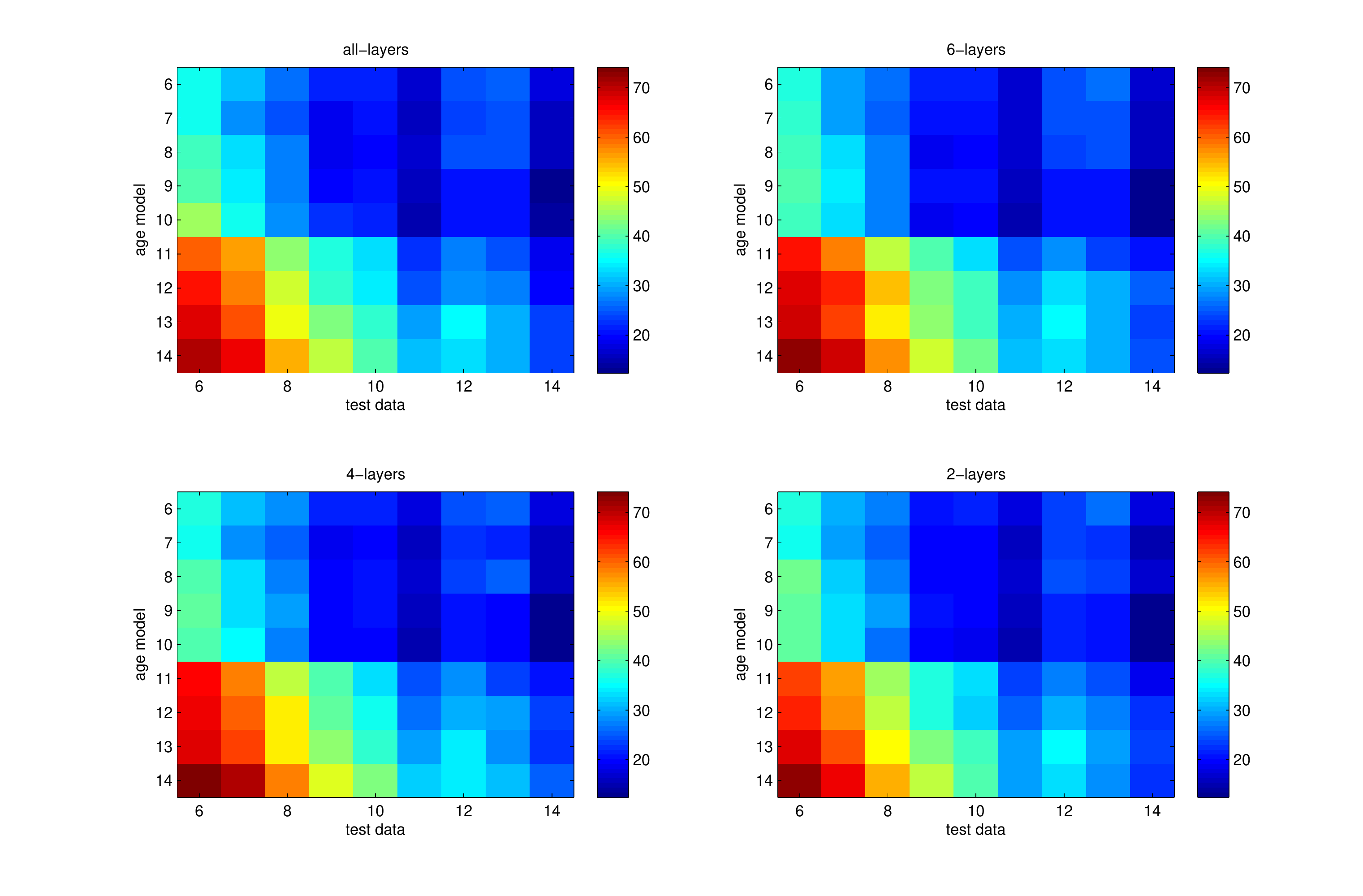}
\caption{Age dependent model performances - All layer configurations}\label{fig:age_dep_confmatrix}
\end{figure*}

Figure~\ref{fig:age_dep_confmatrix} illustrates the confusion matrix obtained by the application of age dependent transformations on each group for each of the 4 adaptation configurations.
A quick inspection shows that all of the configurations exhibit similar trends observed in Section~\ref{sec:age_dependent}.

\section{Age dependent transformations versus Age independent transformations}\label{sec:age_indep_vs_dep}

\begin{figure}[!b]
\centering
\includegraphics[width=0.5\textwidth]{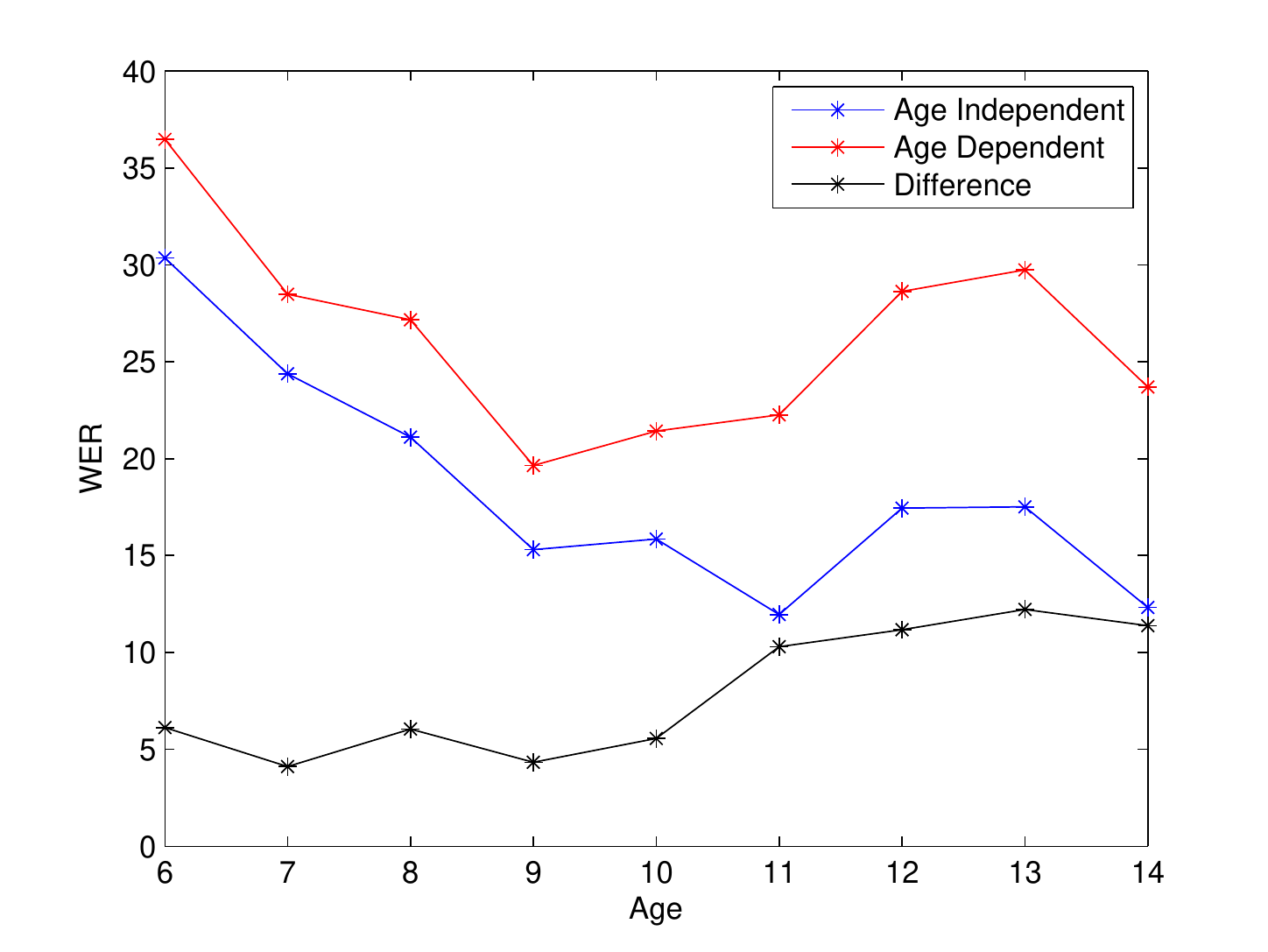}
\caption{Age dependent transformations versus Age independent transformations}\label{fig:age_indep_vs_dep}
\end{figure}

Figure~\ref{fig:age_indep_vs_dep} compares the performance of the age independent transformations (obtained by adapting on all the data) against the application of matched age dependent transformations.
To keep the analysis consistent over different adaptation layer configurations, we consider only the configuration of adapting all the layers.
We find that the age independent transformation trained on significantly more data outperforms the age dependent transformations consistently over all the ages.
This finding suggests that DNN can exploit more data to offset and surpass the performance and effectively generalize over different ages due to its large parameter space.
It does not lose its generalizability when exposed to different ages.
This is in contrast to GMM models, that when adapted (e.g. via MLLR), to a wider diverse population with limited data underperform specific adaptations \cite{gales1996mean}.
By examining the difference between the WER trajectories over age, we find a peak over the ages 11 years to 14 years, highlighting the aforementioned effect of limited data for these age groups as in Figure~\ref{fig:age_duration_distribution}.

\begin{figure*}[t]
\centering
\includegraphics[width=0.75\textwidth,trim={2.5cm 1cm 2.5cm 1cm},clip]{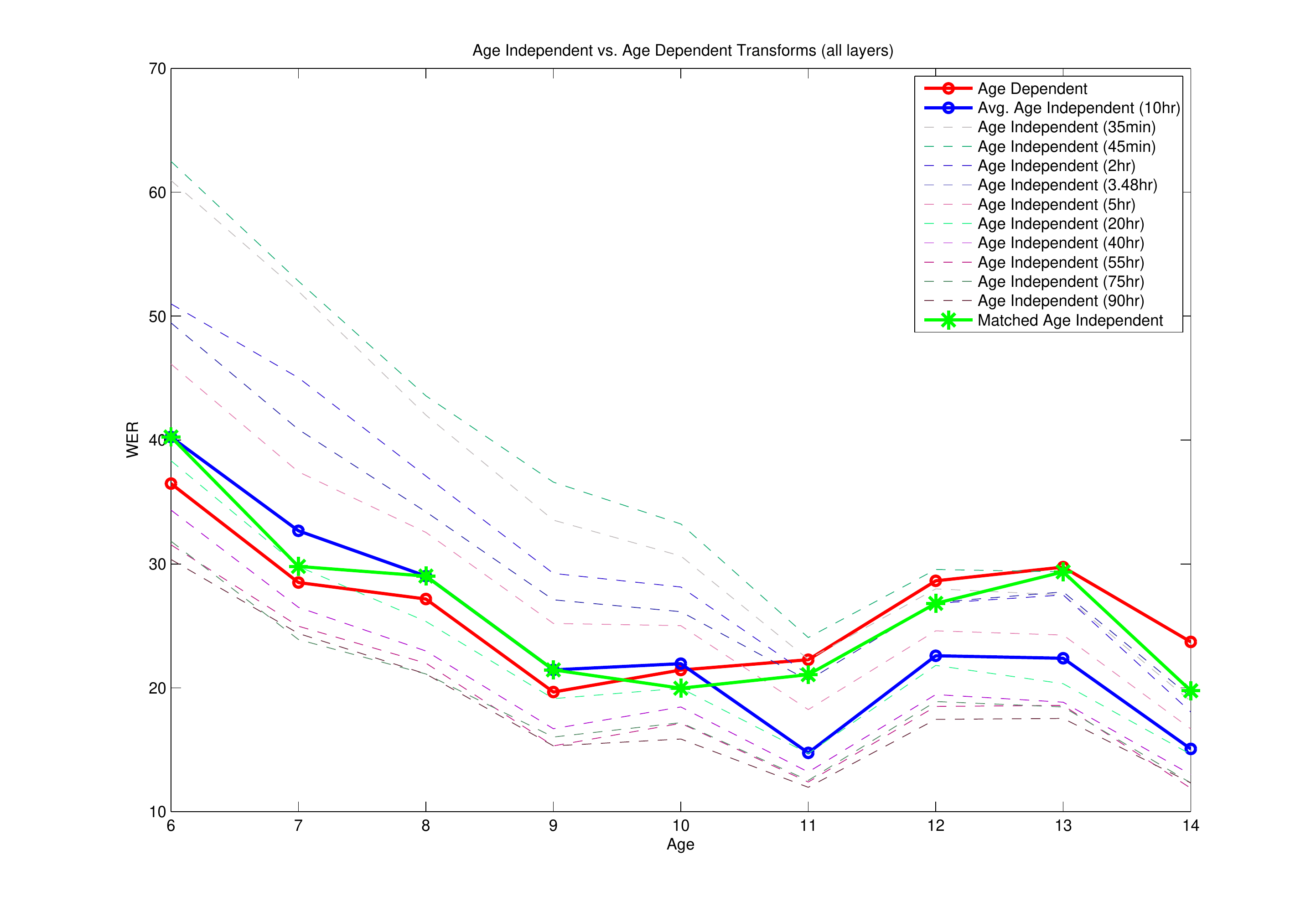}
\caption{Age dependent transformations versus data-normalized Age independent transformations (Adapting all layers)}\label{fig:age_indep_vs_dep_all}
\end{figure*}

However, by providing a correction factor to compensate for the difference between the amount of data between the age dependent and independent transforms, enables for a more fair comparison between the transforms. To enable such an analysis we adopt 2 different types of data correction factors for age independent transformation:


\begin{enumerate}
\item We restrict the amount of data used for the computation of age independent transformation by taking the average data (over ages - Figure~\ref{fig:age_duration_distribution}) which in our case is approximately 10 hours.  We refer to this as average age-independent transform (Blue line in Figure~\ref{fig:age_indep_vs_dep_all}).


\item We train multiple age independent transforms by restricting the amount of adaptation data closest to that of each age. This gives us one age-independent model for each age best matched to age-dependent transform in terms of adaptation data. We refer to this as matched age-independent transform (Green line in Figure~\ref{fig:age_indep_vs_dep_all}).

\end{enumerate}
Since the sampling of data in either case is random, this retains the original corpus proportions (with respect to age).

Figure~\ref{fig:age_indep_vs_dep_all} compares the data normalized age independent transform against the age dependent transformations. The following observations are apparent from the plot:
\begin{enumerate}
\item After normalizing the amount of data, we now see that the age dependent transformations outperform the age independent transformations for younger children (ages 6 years to 10 years) in both cases (average and matched versions). 
\item We observe that the improvements from age dependent transforms gets more prominent as the age decreases, with maximum gains for 6 year old. 
\item We observe a crossover for elder children (ages 11 years to 14 years) in both the cases of average and matched versions, i.e., the age independent transformations are better compared to that of age dependent. (For elder children, the average version of age independent transformation shows higher demarcation due to the heightened mismatch in adaptation data. Hence, the matched version is more representative.). This interesting finding could be attributed towards the higher similarity between the speech of adults and elder children. (Note: this is not a case of age-dependent acoustic modeling, but rather an adaptation from adult's speech). 

\item Looking at the difference between the `best performing' age-independent transformation and the age dependent transform, i.e., the potential gains from exploiting more data with age-independent transform increases with increase in age. This is expected, considering that the elder children exhibit relatively lower variability in acoustic and pronunciation constructs and hence exhibit much similar speech structure to that of the adult.
\end{enumerate}

\subsection{Effect of adaptation layer configurations}

\begin{figure}[!b]
\centering
\includegraphics[width=1.0\columnwidth,trim={0 0 0 1cm}]{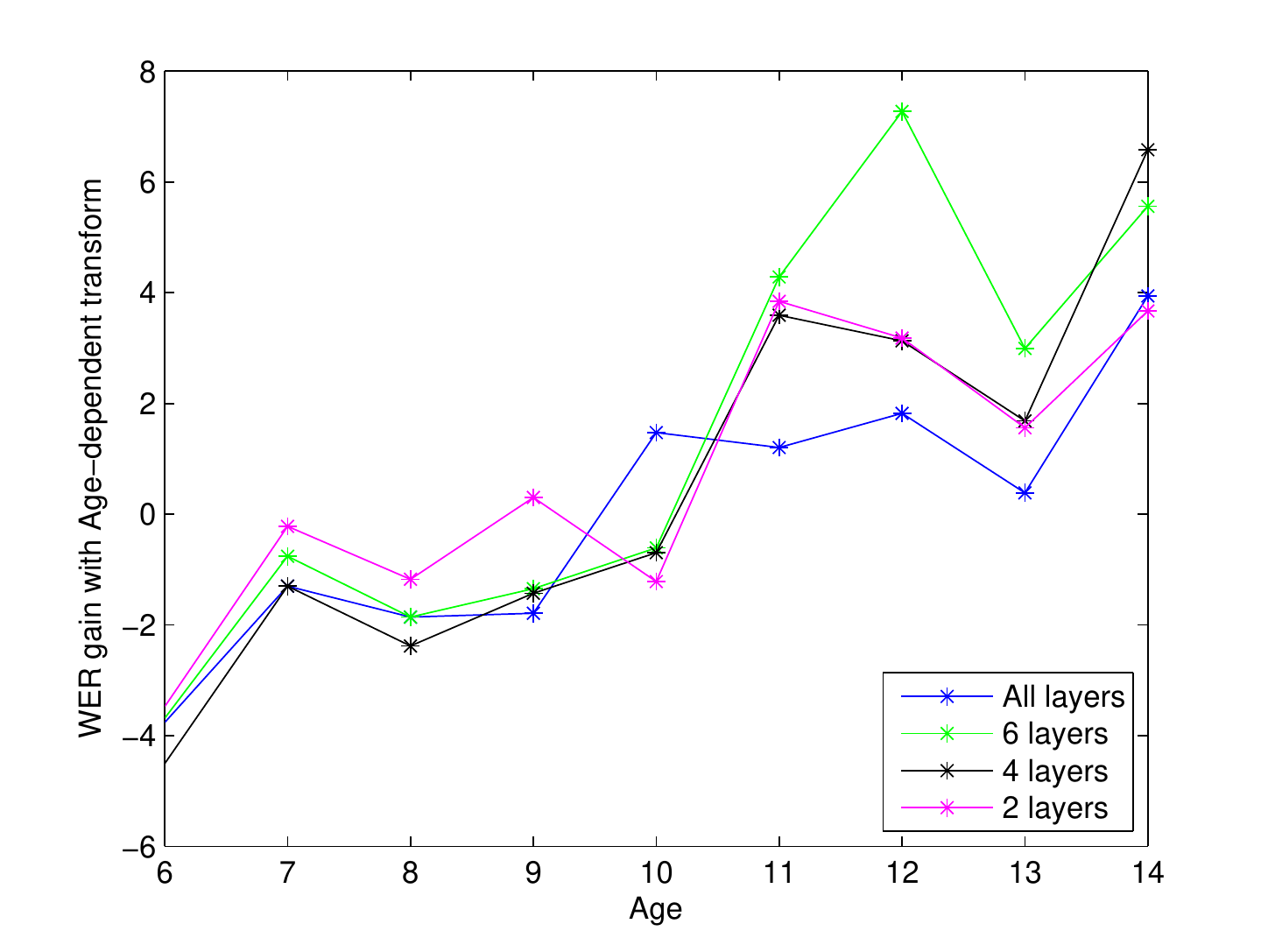}
\caption{Effect of adaptation layer configurations: Difference of WER between Age dependent transformations and matched age-independent transformations}\label{fig:age_indep_vs_dep_layerconfig}
\end{figure}

Figure~\ref{fig:age_indep_vs_dep_layerconfig} plots the difference between the age-dependent transform and the matched version of age-independent transform for different adaptation layer configurations. 
The takeout from the plot is, the age-dependent transforms outperform the age-independent transforms for younger children, whereas the age-independent transforms are beneficial for elder children.
The trend observed earlier, with all the layers, remains apparent over all the layer configurations. 
Note the absolute values (trajectories) are a function of the amount of data present for each age and age itself, as supported by our earlier observations in Section~\ref{sec:layerconfig_data_age}. 
Hence, the inter-relations of different configuration trajectories is complex. 

\section{Conclusion \& Future Work}\label{sec:conclusion}
\Pg{Conclusion}
In this study, we conduct an analysis of LVCSR adaptation and transfer learning for children's speech using multiple databases.
We compare the advantages of DNN acoustic models over the GMM-HMM systems.
We also compare adult and children DNN acoustic model performance for decoding children speech.
Several transfer learning techniques are evaluated, on adult models, specifically to address the increased acoustic variability and pronunciation variability found in children.
Extensive analysis is performed to study the effect of the amount of adaptation data, DNN transfer learning configurations and their impact on different age groups.
In the case of severely limited in-domain (kids) data we proposed and analyzed disjoint adaptation.
We also analyzed the amount of adaptation data required for children of different ages.
We investigated various transfer learning configurations and their effect on different age groups and data sizes.
Our work validated the benefits of age dependent transfer learning and examined the portability and extensibility of models over the different age groups.
We also presented comparisons of age dependent and age independent transfer learning.
These provide valuable insights towards future research directions in terms of persisting challenges and problems in children's speech recognition.

\Pg{Future Work}
In future we would like to analyze the variability internal to the DNN, i.e. how the weights of the ``adapted layers'' change.
Comparisons of such variability between adult and child models can inform on linguistic and structural aspects of kids speech.
These can also help identify the aspects of non-linearities in normalization and adaptation techniques towards improved kids speech processing.
Such models can provide insights in analyzing the effect on various speech parameters in regards to pitch, intensity, voice quality, duration, formant frequencies etc, which are valuable aspects in assessing the difficulties faced for children ASR. 



\section*{Financial Support}
The U.S. Army Medical Research Acquisition Activity, 820 Chandler Street, Fort Detrick MD 21702- 5014 is the awarding and administering acquisition office. This work was supported by the Office of the Assistant Secretary of Defense for Health Affairs through the Psychological Health and Traumatic Brain Injury Research Program under Award No. W81XWH-15-1-0632. Opinions, interpretations, conclusions and recommendations are those of the author and are not necessarily endorsed by the Department of Defense.

\balance
\bibliographystyle{IEEEtran} 
\bibliography{IEEEabrv,refs}

\begin{thebibliography}{10}
\providecommand{\url}[1]{#1}
\csname url@samestyle\endcsname
\providecommand{\newblock}{\relax}
\providecommand{\bibinfo}[2]{#2}
\providecommand{\BIBentrySTDinterwordspacing}{\spaceskip=0pt\relax}
\providecommand{\BIBentryALTinterwordstretchfactor}{4}
\providecommand{\BIBentryALTinterwordspacing}{\spaceskip=\fontdimen2\font plus
\BIBentryALTinterwordstretchfactor\fontdimen3\font minus
  \fontdimen4\font\relax}
\providecommand{\BIBforeignlanguage}[2]{{%
\expandafter\ifx\csname l@#1\endcsname\relax
\typeout{** WARNING: IEEEtran.bst: No hyphenation pattern has been}%
\typeout{** loaded for the language `#1'. Using the pattern for}%
\typeout{** the default language instead.}%
\else
\language=\csname l@#1\endcsname
\fi
#2}}
\providecommand{\BIBdecl}{\relax}
\BIBdecl

\bibitem{potamianos2003robust}
A.~Potamianos and S.~Narayanan, ``Robust recognition of children's speech,''
  \emph{IEEE Transactions on speech and audio processing}, vol.~11, no.~6, pp.
  603--616, 2003.

\bibitem{lee1999acoustics}
S.~Lee, A.~Potamianos, and S.~Narayanan, ``Acoustics of children’s speech:
  Developmental changes of temporal and spectral parameters,'' \emph{The
  Journal of the Acoustical Society of America}, vol. 105, no.~3, pp.
  1455--1468, 1999.

\bibitem{gerosa2006acoustic}
M.~Gerosa, D.~Giuliani, and S.~Narayanan, ``Acoustic analysis and automatic
  recognition of spontaneous children's speech,'' in \emph{Ninth International
  Conference on Spoken Language Processing}, 2006.

\bibitem{potamianos1997automatic}
A.~Potamianos, S.~Narayanan, and S.~Lee, ``Automatic speech recognition for
  children.'' in \emph{Eurospeech}, 1997.

\bibitem{li2002analysis}
Q.~Li and M.~J. Russell, ``An analysis of the causes of increased error rates
  in children's speech recognition,'' in \emph{Seventh International Conference
  on Spoken Language Processing}, 2002.

\bibitem{russell2001automatic}
Q.~L. M.~J. Russell, ``Why is automatic recognition of children's speech
  difficult?'' 2001.

\bibitem{shivakumar2014improving}
P.~G. Shivakumar, A.~Potamianos, S.~Lee, and S.~Narayanan, ``Improving speech
  recognition for children using acoustic adaptation and pronunciation
  modeling,'' in \emph{Proc. Workshop on Child, Computer and Interaction
  (WOCCI)}, 2014.

\bibitem{umesh2007study}
S.~Umesh and R.~Sinha, ``A study of filter bank smoothing in mfcc features for
  recognition of children's speech,'' \emph{IEEE Transactions on audio, speech,
  and language processing}, vol.~15, no.~8, pp. 2418--2430, 2007.

\bibitem{ghai2015pitch}
S.~Ghai and R.~Sinha, ``Pitch adaptive mfcc features for improving children’s
  mismatched asr,'' \emph{International Journal of Speech Technology}, vol.~18,
  no.~3, pp. 489--503, 2015.

\bibitem{shahnawazuddin2016pitch}
S.~Shahnawazuddin, A.~Dey, and R.~Sinha, ``Pitch-adaptive front-end features
  for robust children's asr.'' in \emph{INTERSPEECH}, 2016, pp. 3459--3463.

\bibitem{giuliani2003investigating}
D.~Giuliani and M.~Gerosa, ``Investigating recognition of children's speech,''
  in \emph{Acoustics, Speech, and Signal Processing, 2003.
  Proceedings.(ICASSP'03). 2003 IEEE International Conference on},
  vol.~2.\hskip 1em plus 0.5em minus 0.4em\relax IEEE, 2003, pp. II--137.

\bibitem{stemmer2003acoustic}
G.~Stemmer, C.~Hacker, S.~Steidl, and E.~N{\"o}th, ``Acoustic normalization of
  children's speech.'' in \emph{INTERSPEECH}, 2003.

\bibitem{elenius2005adaptation}
D.~Elenius and M.~Blomberg, ``Adaptation and normalization experiments in
  speech recognition for 4 to 8 year old children.'' in \emph{Interspeech},
  2005, pp. 2749--2752.

\bibitem{gray2014child}
S.~S. Gray, D.~Willett, J.~Lu, J.~Pinto, P.~Maergner, and N.~Bodenstab, ``Child
  automatic speech recognition for us english: child interaction with
  living-room-electronic-devices,'' in \emph{Proceedings of workshop on child
  computer interaction (WOCCI)}, 2014.

\bibitem{gerosa2009review}
M.~Gerosa, D.~Giuliani, S.~Narayanan, and A.~Potamianos, ``A review of asr
  technologies for children's speech,'' in \emph{Proceedings of the 2nd
  Workshop on Child, Computer and Interaction}.\hskip 1em plus 0.5em minus
  0.4em\relax ACM, 2009, p.~7.

\bibitem{potamianos1998spoken}
A.~Potamianos and S.~Narayanan, ``Spoken dialog systems for children,'' in
  \emph{Acoustics, Speech and Signal Processing, 1998. Proceedings of the 1998
  IEEE International Conference on}, vol.~1.\hskip 1em plus 0.5em minus
  0.4em\relax IEEE, 1998, pp. 197--200.

\bibitem{das1998improvements}
S.~Das, D.~Nix, and M.~Picheny, ``Improvements in children's speech recognition
  performance,'' in \emph{Acoustics, Speech and Signal Processing, 1998.
  Proceedings of the 1998 IEEE International Conference on}, vol.~1.\hskip 1em
  plus 0.5em minus 0.4em\relax IEEE, 1998, pp. 433--436.

\bibitem{fringi2015evidence}
E.~Fringi, J.~F. Lehman, and M.~Russell, ``Evidence of phonological processes
  in automatic recognition of children's speech,'' in \emph{Sixteenth Annual
  Conference of the International Speech Communication Association}, 2015.

\bibitem{Tepperman2011Agenerativestudentmodel}
J.~Tepperman, S.~Lee, S.~S. Narayanan, and A.~Alwan, ``A generative student
  model for scoring word reading skills,'' \emph{IEEE Transactions on Audio,
  Speech, and Language Processing}, vol.~19, no.~2, pp. 348 -- 360, 2011.

\bibitem{tulsiani2017acoustic}
H.~Tulsiani, P.~Swarup, and P.~Rao, ``Acoustic and language modeling for
  children's read speech assessment,'' in \emph{Communications (NCC), 2017
  Twenty-third National Conference on}.\hskip 1em plus 0.5em minus 0.4em\relax
  IEEE, 2017, pp. 1--6.

\bibitem{tong2017multi}
R.~Tong, N.~F. Chen, and B.~Ma, ``Multi-task learning for mispronunciation
  detection on singapore children’s mandarin speech,'' \emph{Proc.
  Interspeech 2017}, pp. 2193--2197, 2017.

\bibitem{hagen2007highly}
A.~Hagen, B.~Pellom, and R.~Cole, ``Highly accurate children’s speech
  recognition for interactive reading tutors using subword units,''
  \emph{speech communication}, vol.~49, no.~12, pp. 861--873, 2007.

\bibitem{giuliani2015large}
D.~Giuliani and B.~BabaAli, ``Large vocabulary children's speech recognition
  with dnn-hmm and sgmm acoustic modeling,'' in \emph{Sixteenth Annual
  Conference of the International Speech Communication Association}, 2015.

\bibitem{cosi2015kaldi}
P.~Cosi, ``A kaldi-dnn-based asr system for italian,'' in \emph{2015
  International Joint Conference on Neural Networks (IJCNN)}.\hskip 1em plus
  0.5em minus 0.4em\relax IEEE, 2015, pp. 1--5.

\bibitem{serizel2014vocal}
R.~Serizel and D.~Giuliani, ``Vocal tract length normalisation approaches to
  dnn-based children's and adults' speech recognition,'' in \emph{Spoken
  Language Technology Workshop (SLT), 2014 IEEE}.\hskip 1em plus 0.5em minus
  0.4em\relax IEEE, 2014, pp. 135--140.

\bibitem{liao2015large}
H.~Liao, G.~Pundak, O.~Siohan, M.~Carroll, N.~Coccaro, Q.-M. Jiang, T.~N.
  Sainath, A.~Senior, F.~Beaufays, and M.~Bacchiani, ``Large vocabulary
  automatic speech recognition for children,'' 2015.

\bibitem{qian2016mismatched}
M.~Qian, I.~McLoughlin, W.~Quo, and L.~Dai, ``Mismatched training data
  enhancement for automatic recognition of children's speech using dnn-hmm,''
  in \emph{Chinese Spoken Language Processing (ISCSLP), 2016 10th International
  Symposium on}.\hskip 1em plus 0.5em minus 0.4em\relax IEEE, 2016, pp. 1--5.

\bibitem{fainberg2016improving}
J.~Fainberg, P.~Bell, M.~Lincoln, and S.~Renals, ``Improving children's speech
  recognition through out-of-domain data augmentation.'' in \emph{INTERSPEECH},
  2016, pp. 1598--1602.

\bibitem{qian2016improving}
Y.~Qian, X.~Wang, K.~Evanini, and D.~Suendermann-Oeft, ``Improving dnn-based
  automatic recognition of non-native children speech with adult speech,'' in
  \emph{Workshop on Child Computer Interaction}, 2017, pp. 40--44.

\bibitem{tong2017transfer}
R.~Tong, L.~Wang, and B.~Ma, ``Transfer learning for children's speech
  recognition,'' in \emph{Asian Language Processing (IALP), 2017 International
  Conference on}.\hskip 1em plus 0.5em minus 0.4em\relax IEEE, 2017, pp.
  36--39.

\bibitem{serizel2014deep}
R.~Serizel and D.~Giuliani, ``Deep neural network adaptation for children’s
  and adults’ speech recognition,'' in \emph{Proc. of the First Italian
  Computational Linguistics Conference}, 2014.

\bibitem{serizel2017deep}
------, ``Deep-neural network approaches for speech recognition with
  heterogeneous groups of speakers including children,'' \emph{Natural Language
  Engineering}, vol.~23, no.~3, pp. 325--350, 2017.

\bibitem{matassoni2018}
M.~Matassoni, R.~Gretter, and G.~D. Falavigna, Daniele~and, ``Non-native
  children speech recognition through transfer learning,'' \emph{Acoustics,
  Speech and Signal Processing (ICASSP), 2018 IEEE International Conference
  on}, pp. 6229--6233, apr 2018.

\bibitem{heigold2013multilingual}
G.~Heigold, V.~Vanhoucke, A.~Senior, P.~Nguyen, M.~Ranzato, M.~Devin, and
  J.~Dean, ``Multilingual acoustic models using distributed deep neural
  networks,'' in \emph{2013 IEEE International Conference on Acoustics, Speech
  and Signal Processing}.\hskip 1em plus 0.5em minus 0.4em\relax IEEE, 2013,
  pp. 8619--8623.

\bibitem{huang2013cross}
J.-T. Huang, J.~Li, D.~Yu, L.~Deng, and Y.~Gong, ``Cross-language knowledge
  transfer using multilingual deep neural network with shared hidden layers,''
  in \emph{2013 IEEE International Conference on Acoustics, Speech and Signal
  Processing}.\hskip 1em plus 0.5em minus 0.4em\relax IEEE, 2013, pp.
  7304--7308.

\bibitem{cirecsan2012transfer}
D.~C. Cire{\c{s}}an, U.~Meier, and J.~Schmidhuber, ``Transfer learning for
  latin and chinese characters with deep neural networks,'' in \emph{The 2012
  International Joint Conference on Neural Networks (IJCNN)}.\hskip 1em plus
  0.5em minus 0.4em\relax IEEE, 2012, pp. 1--6.

\bibitem{bengio2013representation}
Y.~Bengio, A.~Courville, and P.~Vincent, ``Representation learning: A review
  and new perspectives,'' \emph{IEEE transactions on pattern analysis and
  machine intelligence}, vol.~35, no.~8, pp. 1798--1828, 2013.

\bibitem{dehak2011front}
N.~Dehak, P.~Kenny, R.~Dehak, P.~Dumouchel, and P.~Ouellet, ``Front-end factor
  analysis for speaker verification,'' \emph{Audio, Speech, and Language
  Processing, IEEE Transactions on}, vol.~19, no.~4, pp. 788--798, 2011.

\bibitem{shivakumar2014simplified}
P.~G. Shivakumar, M.~Li, V.~Dhandhania, and S.~S. Narayanan, ``Simplified and
  supervised i-vector modeling for speaker age regression,'' in
  \emph{Acoustics, Speech and Signal Processing (ICASSP), 2014 IEEE
  International Conference on}.\hskip 1em plus 0.5em minus 0.4em\relax IEEE,
  2014, pp. 4833--4837.

\bibitem{saon2013speaker}
G.~Saon, H.~Soltau, D.~Nahamoo, and M.~Picheny, ``Speaker adaptation of neural
  network acoustic models using i-vectors.'' in \emph{ASRU}, 2013, pp. 55--59.

\bibitem{cole2006university2}
R.~Cole, P.~Hosom, and B.~Pellom, ``University of colorado prompted and read
  children’s speech corpus,'' Technical Report TR-CSLR-2006-02, Center for
  Spoken Language Research, University of Colorado, Boulder, Tech. Rep., 2006.

\bibitem{cole2006university}
R.~Cole and B.~Pellom, ``University of colorado read and summarized story
  corpus,'' Technical Report TR-CSLR-2006-03, University of Colorado, Tech.
  Rep., 2006.

\bibitem{shobaki2000ogi}
K.~Shobaki, J.-P. Hosom, and R.~Cole, ``The ogi kids' speech corpus and
  recognizers,'' in \emph{Proc. of ICSLP}, 2000, pp. 564--567.

\bibitem{rousseau2014enhancing}
A.~Rousseau, P.~Del{\'e}glise, and Y.~Est{\`e}ve, ``Enhancing the ted-lium
  corpus with selected data for language modeling and more ted talks.'' in
  \emph{LREC}, 2014, pp. 3935--3939.

\bibitem{weide1998cmu}
R.~Weide, ``The cmu pronunciation dictionary, release 0.6,'' 1998.

\bibitem{walker2004sphinx}
W.~Walker, P.~Lamere, P.~Kwok, B.~Raj, R.~Singh, E.~Gouvea, P.~Wolf, and
  J.~Woelfel, ``Sphinx-4: A flexible open source framework for speech
  recognition,'' 2004.

\bibitem{peddinti2015time}
V.~Peddinti, D.~Povey, and S.~Khudanpur, ``A time delay neural network
  architecture for efficient modeling of long temporal contexts,'' in
  \emph{Sixteenth Annual Conference of the International Speech Communication
  Association}, 2015.

\bibitem{bengio2007greedy}
Y.~Bengio, P.~Lamblin, D.~Popovici, H.~Larochelle \emph{et~al.}, ``Greedy
  layer-wise training of deep networks,'' \emph{Advances in neural information
  processing systems}, vol.~19, p. 153, 2007.

\bibitem{gales1996mean}
M.~J. Gales and P.~C. Woodland, ``Mean and variance adaptation within the mllr
  framework,'' \emph{Computer Speech \& Language}, vol.~10, no.~4, pp.
  249--264, 1996.

\end{thebibliography}





\end{document}